\documentclass[12pt,preprint]{aastex}
\usepackage{natbib}
\usepackage{amsmath,amsfonts,amssymb}
\usepackage{graphicx}
\usepackage{multirow}
\usepackage{color}
\def\url#1{\texttt{#1}}

\newcommand{\psini}{\ensuremath{P \left(\sin i_{\star}\right)^{-1}}}
\newcommand{\vrotsini}{\ensuremath{v_\mathrm{rot}\sin i_{\star}}}
\newcommand{\mps}{\thinspace\ensuremath{\mathrm{m\,s}^{-1}}}
\newcommand{\kmps}{\thinspace\ensuremath{\mathrm{km\,s^{-1}}}}
\newcommand{\Teff}{\ensuremath{T_\mathrm{eff}}}
\newcommand{\Msun}{\thinspace\ensuremath{\mathrm{M}_\odot}}
\newcommand{\Mjup}{\thinspace\ensuremath{\mathrm{M}_\mathrm{J}}}
\newcommand{\Rsun}{\thinspace\ensuremath{\mathrm{R}_\odot}}

\newcommand{\llsun}{\hbox{$\log L/L_{\odot}$}}
\shorttitle{Three red giants with planetary candidates}
\shortauthors{Niedzielski et al.}

\begin{document}

\title{Three red giants with substellar-mass companions\footnote{Based on observations obtained with the Hobby-Eberly Telescope, which is a joint project of the University of Texas at Austin, the Pennsylvania State University, Stanford University, Ludwig-Maximilians-Universität München, and Georg-August-Universität Göttingen.}. }
 \author{ 
  A. Niedzielski,$^1$ 
  A. Wolszczan,$^{2,3}$
           G. Nowak,$^{4,5,1}$
  M. Adam\'ow,$^{6,1}$
             K. Kowalik,$^{7,1}$
                      G. Maciejewski,$^1$
       B. Deka-Szymankiewicz,$^1$
       M. Adamczyk$^1$
  }

  \altaffiltext{1}{Toru\'n Centre for Astronomy, Nicolaus Copernicus University in Toru\'n, Grudziadzka 5, 87-100 Toru\'n, Poland, \email{Andrzej.Niedzielski@umk.pl}}
 \altaffiltext{2}{Department of Astronomy and Astrophysics, Pennsylvania State University, 525 Davey Laboratory, University Park, PA 16802, USA, \email{alex@astro.psu.edu}}
\altaffiltext{3}{Center for Exoplanets and Habitable Worlds, Pennsylvania State University, 525 Davey Laboratory, University Park, PA 16802, USA}
\altaffiltext{4}{Instituto de Astrof\'{\i}sica de Canarias, C/ V\'{\i}a L\'actea, s/n, E38205 - La Laguna (Tenerife), Spain}
\altaffiltext{5}{Departamento de Astrof\'{\i}sica, Universidad de La Laguna, E-38206 La Laguna, Tenerife, Spain.}
\altaffiltext{6}{McDonald Observatory and Department of Astronomy, University of Texas at Austin, 2515 Speedway, Stop C1402, Austin, Texas, 78712-1206, USA.}
\altaffiltext{7}{National Center for Supercomputing Applications, University of Illinois, Urbana-Champaign, 1205 W Clark St, MC-257, Urbana, IL 61801, USA}

\begin{abstract}

We present  three giant stars from the ongoing Penn State-Toru\'n Planet Search  with the Hobby-Eberly Telescope, 
which exhibit radial velocity variations that point to a presence of planetary --mass companions around them. 
BD+49~828 is a $M=1.52 \pm 0.22 \Msun$ K0 giant with a $m \sin i=1.6^{+0.4}_{-0.2} \Mjup$  minimum mass 
companion in $a=4.2^{+0.32}_{-0.2}$~AU ($2590^{+300}_{-180}$d), $e=0.35^{+0.24}_{-0.10}$ orbit.
HD~95127, a  $\llsun=2.28 \pm 0.38$, $R = 20\pm 9 \Rsun$, $M=1.20 \pm0.22 \Msun$ K0 giant 
has a m{\it{sini}}=$5.01^{+0.61}_{-0.44}$ \Mjup   minimum mass companion 
in $a=1.28^{+0.01}_{-0.01}$~AU ($482^{+5}_{-5}$d), $e=0.11^{+0.15}_{-0.06}$ orbit.
Finally, HD 216536, is a  $M=1.36 \pm 0.38 \Msun$ K0 giant with a $m \sin i=1.47^{+0.20}_{-0.12} \Mjup$ minimum 
mass companion  in $a=0.609^{+0.002}_{-0.002}$~AU ($148.6^{+0.7}_{-0.7}$d), $e=0.38^{+0.12}_{-0.10}$ orbit.
Both, HD~95127~b  and HD~216536~b 
in their   compact orbits, are very close to the engulfment zone and hence prone to ingestion in the near future.
BD+49~828~b 
is among the longest period planets detected with the radial velocity technique until now and it will remain 
unaffected by stellar evolution up to a very late stage of its host. We discuss general properties  of planetary systems around evolved stars 
and planet survivability using existing data on exoplanets in more detail.

\end{abstract}

\keywords{planetary systems-stars: individual (BD+49~828 b;   HD~95127 b; HD~216536 b)}

\section{Introduction}

Searches for planets around stars that evolved off the Main Sequence (MS) represent an important 
ingredient of the exoplanet research. They extend the known population of planets to those orbiting 
massive and intermediate-mass stars. They also provide the best tool  to study star-planet interactions 
induced by stellar evolution.  Finally, these searches furnish high quality spectroscopic data that can be used 
in a wide range of studies related to stellar evolution.

Intermediate-mass and massive MS stars have high effective temperatures and rotate rapidly.  With only a few,
rotation broadened spectral lines, these stars are not suitable for high precision radial velocity (RV) searches for planetary companions.
Consequently,  planets around intermediate-mass and massive MS stars discovered occasionally 
in transit searches (cf. \citealt{2011ApJ...736...19B, 2013ApJ...768..127S}) are very difficult to confirm with RV measurements. 
The main source of data on such planetary systems are RV searches that focus on  subgiant and giant stars which are
evolving off the MS, cooling down, and considerably slowing their rotation. The most massive hosts of planetary systems 
come almost exclusively from such surveys  (e.g. \citealt{ 2007ApJ...661..527S, 2010PASJ...62.1063S, 2012PASJ...64..135S, 2013PASJ...65...85S}).
Surveys of this kind have been uncovering a population of planetary systems that are descendants of an undetermined 
original sample of planets around intermediate-mass and massive stars, carrying with them imprints of the effects 
of several Gyr of dynamical evolution enforced by an evolving star. They have demonstrated, for example, a paucity of planets 
within 0.5~AU of their parent stars \citep{2007ApJ...665..785J, 2008PASJ...60..539S}, a borderline currently 
broken by Kepler 91b, a planet  at only $a/R_{\star}=2.46$ \citep{2014A&A...562A.109L, 2014arXiv1408.3149B}.

The other frequently discussed effects
are the higher frequency of planets around more massive stars and the stellar mass - planetary 
system mass relation \citep{2007A&A...472..657L, 2010ApJ...721L.153J, 2011ApJS..197...26J}. Such stars also seem to prefer 
massive, $m_{p} \geq 1.5 \Mjup$, planets \citep{2007ApJ...661..527S, 2009ApJ...707..768N, 2010ApJ...709..396B}  and a more 
frequent occurrence of brown dwarf - mass companions, which are deficient around solar-mass stars 
on the MS \citep{2008ApJ...672..553L, 2009ApJ...707..768N, 2013A&A...555A..87M, 2013MmSAI..84.1035N}.

 Disentangling information on the original planet population from the available data on  planets around giants requires a detailed understanding of the 
 interaction between an orbiting planet and an evolving star.
 In fact, importance of the influence of tidal interactions on orbital evolution has been recognized long before first exoplanets were discovered 
 (see \citealt{1972Ap&SS..16....3K, 1973ApJ...180..307C, 1977A&A....57..383Z} and references therein). \cite{1982A&A...112..190L} studied 
 the net effect of accretion drag and mass loss in a star-planet system.
 The first attempts to predict  planet survival upon stellar evolution were presented by \cite{1980PhDT.........4C, 1981Ap&SS..77..319C}, 
 \cite{1983A&A...125L..12L, 1984MNRAS.208..763L} and \cite{1984EM&P...31..313V}. 
 Detailed effects of brown dwarf or planet capture by an evolving star were numerically modeled in  
\cite{1999MNRAS.308.1133S,1999MNRAS.304..925S}. 
The future of the Solar System in the context of the Sun's evolution was studied by \cite{1998Icar..134..303D, 2001Icar..151..130R,2008MNRAS.386..155S}.

A more general picture of stellar evolution induced star-planet interaction, including tidal interactions  was presented 
in \cite{{1996ApJ...470.1187R}, 2007ApJ...661.1192V, {2009ApJ...700..832C},  2009ApJ...698.1357J, 2010MNRAS.408..631N}.
Three possible outcomes of orbital evolution of planets affected by tidal forces and stellar evolution, somewhat analogous 
to those in planet-satellite system \citep{1973ApJ...180..307C} were presented in \cite{2009ApJ...705L..81V}.
 Most recent studies include such effects as evolution of  the primary's spin  \citep{2013MNRAS.432..500N}   or interactions 
 with the Galactic environment \citep{2014MNRAS.437.1127V}. The influence of stellar wind and tidal force prescriptions 
 on planetary orbit evolution has been recently illustrated in \cite{Villaver2014}.
 
In this paper we present three planetary-mass companions around solar-mass stars
from the ongoing Penn State-Toru\'n Planet Search  (PTPS) with the Hobby-Eberly Telescope \citep{2007ApJ...669.1354N, 2008IAUS..249...43N}: 
BD+49 828, HD~95127  and HD~216536, and discuss their fate through their host evolution. We also summarize available data 
on planetary systems of evolved stars and briefly discuss properties of planetary systems at various stages of their hosts  stellar evolution.

The plan of the paper is as follows: in Section \ref{obs_params_rvs} we  describe the observations and present our targets 
and their basic parameters, the measurements and the analysis of RVs. In Section \ref{keplerian_analysis}  we present 
the results of the Keplerian modeling of RV variations, while other sources of stellar variability are considered in Section \ref{activity}. 
Finally, in Section \ref{conclusions} we discuss our results, 
the fate of planets orbiting giant stars, and some additional problems related to
 evolved planetary systems.

\section{Observations, stellar parameters and radial velocitiy measurements}\label{obs_params_rvs}

\subsection{Observations}

Spectroscopic observations presented here were made  with the Hobby-Eberly Telescope (HET) \citep{Ramsey1998} and its High Resolution Spectrograph (HRS) \citep{Tull1998} in the queue scheduled mode \citep{Shetrone2007}. 
The spectrograph was used in the R=60,000 resolution  mode with a I$_2$ gas cell inserted into the optical path, and it was fed with a 2 arcsec fiber. 
The configuration and observing procedure employed in our program were practically identical to those described by \cite{2004ApJ...611L.133C}. 

BD+49~828 (AG+49 353) is a K0 giant  \citep{1997yCat.1061....0D}  with $B=9.43\pm0.03$ \citep{2000A&A...355L..27H} and  $V=9.38\pm0.02$. 
We collected 42 epochs of precise RVs for this star over 3134 days (8.58 yr), between  MJD 53225 and 56359.  
Typical signal-to-noise ratio (S/N) of about 200 was reached in  540-1600~s depending on atmospheric conditions.

RVs of HD~95127 (BD+44 2038), a K0 giant \citep{hd}  with  $B=10.825\pm 0.045$ and $V=8.15\pm0.01$  \citep{2000A&A...355L..27H}  
 were measured at 41 epochs over the period of 2929 days (8 yr) between MJD 53024 and 55953.The S/Ns of the combined star and iodine spectra ranged from 80 to 250. The S/N of the stellar template was 480. 
The exposure time ranged between 170 and 800~s, depending on the atmospheric conditions. 
Hipparcos parallax from \cite{2007A&A...474..653V} puts this star at a distance of $332 \pm110$~pc.

HD~216536 (BD+43 4329 ) is a K0 giant \citep{hd} with $B=10.42\pm0.03$ and $V=9.23\pm0.02$  \citep{2000A&A...355L..27H}. 
The star was observed over 2756 days (7.54 yr),  between MJD 53544 and 56300. 
The typical S/N of  250 was reached in 455 - 1200~s exposures depending on actual conditions. 

\subsection{Stellar parameters}

All the three giants discussed here belong to the PTPS Red Giant Clump sample. They have been studied in detail by \cite{Zielinski2012}, who estimated their atmospheric parameters 
by means of the method of \citet{Takeda2005a, Takeda2005b} and intrinsic colors.  
In a  more detailed chemical abundances analysis of \cite{2014A&A...569A..55A}  with SME \citep{1996A&AS..118..595V} 
no anomalies were detected and new rotation velocities presented.

Luminosity of HD~95127  
was calculated from the Hipparcos parallax and we found it to be surprisingly high for a $\log g=3.82\pm0.27$ and $\Teff=4807\pm90$~K star \citep{Zielinski2012}. 
To clarify that mismatch we reanalysed all available HET/HRS spectra with no iodine imprinted and performed spectroscopic analysis with the method of \citet{Takeda2005a, Takeda2005b} again. 
Resulting, significantly different atmospheric parameters are presented in Table \ref{Tab:Table0}. 

Using these parameters the mass and age were derived  with the Bayesian method based on \cite{2005A&A...436..127J} 
formalism and modified by  \cite{2006A&A...458..609D}, in which we used  theoretical isochrones of \cite{2012MNRAS.427..127B}.
For the other two stars luminosity  was obtained, together with the mass and age, applying the Bayesian approach described above.

Stellar radii were estimated  either from the derived masses and $\log g$  values obtained  from the spectroscopic analysis 
or using derived luminosities and effective temperatures.
 
With the rotational velocities from  \cite{2014A&A...569A..55A} and stellar radii, masses and luminosities derived here we have obtained estimates 
of  maximum rotation periods, as well as periods and amplitudes of $p$-mode oscillations from the scaling relations of \cite{KjeldsenBedding1995}.
 All the information on targets is presented in Table \ref{Tab:Table0}. 
 
\begin{figure}
\centerline{\includegraphics[angle=0,scale=0.5]{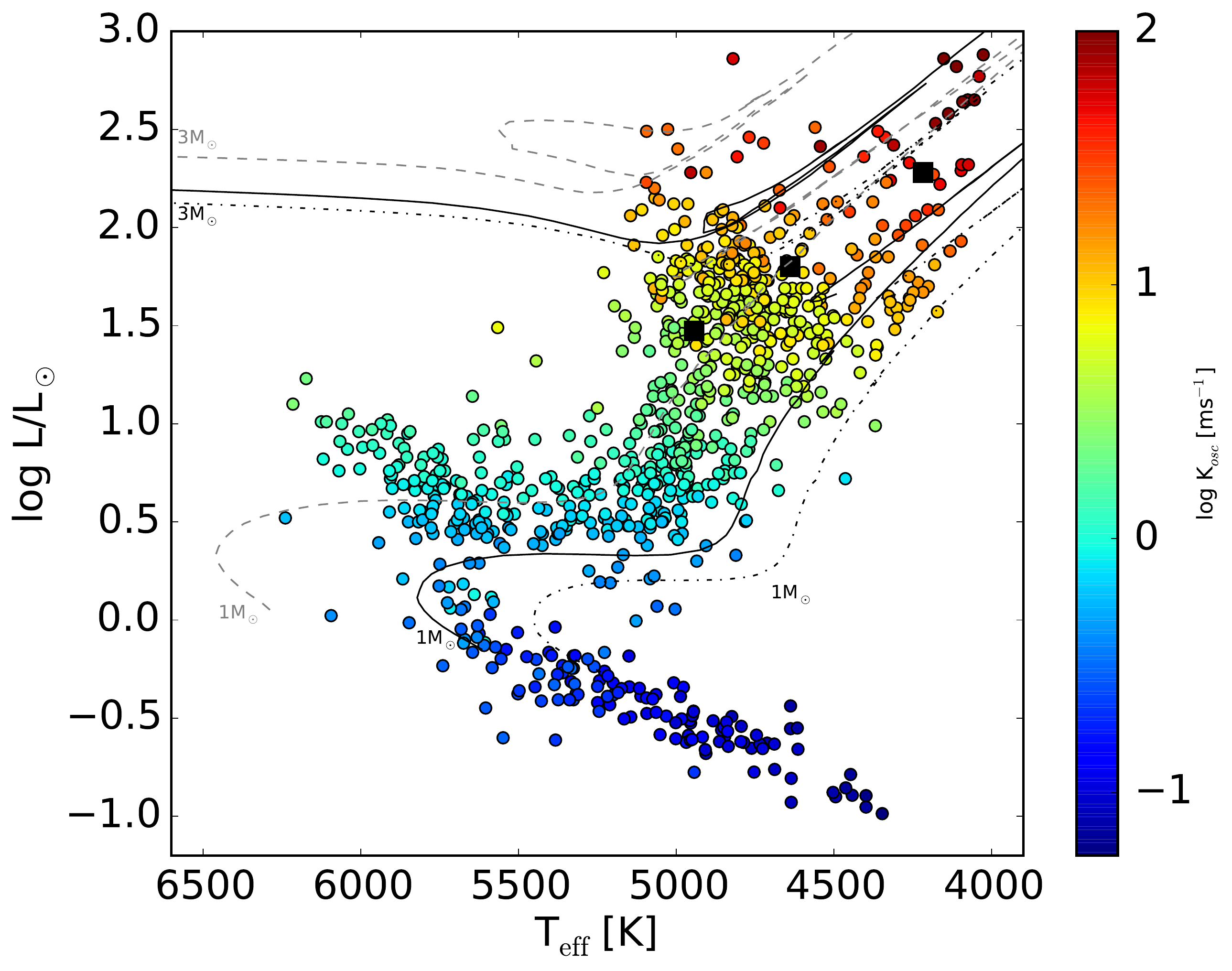}}
\caption{The PTPS sample. Presented are 838 stars with  detailed spectroscopic parameters  available (\citealt{Zielinski2012}; Niedzielski el al. in prep., Deka et al.  in prep.).  The three stars presented in this paper are marked as black rectangles. Color coded is the amplitude of p-mode oscillations estimated from the scaling relations of \cite{KjeldsenBedding1995}. Evolutionary tracks of \cite{2008A&A...484..815B} are presented for three metallicities: [Fe/H]=0.0 (solid line), -0.30 (dashed line) and 0.3 (dotted line)   for illustration purposes.}
\label{figHRD}
\end{figure}

Positions of the three stars discussed in the paper  on 
the Hertzsprung-Russel Diagram (HRD)  are presented in Figure \ref{figHRD}  together with the entire PTPS sample. In this Figure we show 332 stars from \cite{Zielinski2012}, 368 stars from Niedzielski et al. (in prep.) and 138 stars from Deka et al. (in prep.) for which spectroscopically determined atmospheric parameters as well as masses, radii and luminosities are available.

\subsection{Radial velocity measurements.}

The HRS is a general purpose spectrograph which is neither temperature nor pressure controlled and the calibration of RV 
measurements with this instrument is best accomplished with the I$_2$ cell technique. For the purpose of measuring the RVs 
and line bisector velocity span (BS), we have combined the method described by \citep{1992PASP..104..270M, Butler1996} with that 
of cross-correlation used by \citep{1995IAUS..167..221Q, 2002A&A...388..632P}. A detailed implementation of our approach is described 
in \cite{Nowak2012} and \cite{ Nowak2013}. The typical  precision of a few $\mps$  achieved that way made it quite sufficient 
to use the \cite{Stumpff1980} algorithm to refer the measured RVs to the Solar System barycenter.

The precision of both RVs and BSs as well as the long-term stability of our RV measurements has been verified by analyzing the data 
derived from monitoring of stars that do not exhibit detectable RV variations and stars for which the RV or BS variations have been well established. 
One example of such a calibrator, discussed in \cite{2009ApJ...707..768N}, is the K0 giant BD+70 1068 , which has $\sigma_{RV}=12 \mps$ consisting 
of the intrinsic RV uncertainty of  $7 \mps$ and  the approximate $10 \mps$ RV amplitude variation due to solar-type oscillations \cite{KjeldsenBedding1995}. 
\cite{Nowak2013} have carried out a more detailed analysis of our RV precision and stability using the RV measurements 
and orbital solutions for HD 209458, HD 88133. The precision of our BS measurements has been demonstrated for HD~166435 by the same authors.

\section{Keplerian analysis}\label{keplerian_analysis}

Orbital parameters for the companions to the three stars were derived using a hybrid approach (e.g. \citep{{2003ApJ...594.1019G, 2006A&A...449.1219G, 2007ApJ...657..546G}}), 
which combines the PIKAIA-based, global genetic algorithm (GA; \cite{1995ApJS..101..309C})  with a faster and more precise local method. 
Given a periodic signal found in the RV data with the Lomb-Scargle (LS) periodogram  \citep{{1976Ap&SS..39..447L, 1982ApJ...263..835S, 1992nrfa.book.....P}},
we begin with launching a search for possible orbital solutions over a wide range of parameters using the GA code.
The GA semi-global search usually establishes a narrow parameter range that defines the search space for the next step, 
in which the MPFit algorithm \citep{2009ASPC..411..251M} is used to find the best-fit Keplerian orbit  delivered by RVLIN \citep{2009ApJS..182..205W}
modified  to allow the stellar jitter to be fitted as a free parameter \citep{2007ASPC..371..189F, 2011ApJS..197...26J}.  
The RV bootstrapping  method  \citep{1993ApJ...413..349M, 1997A&A...320..831K, 2005ApJ...619..570M, 2007ApJ...657..533W}  
is employed to assess the uncertainties of the best-fit orbital parameters  defined as  the width of the resulting distribution of $10^6$ trials 
of scrambled data between the 15.87th and 84.13th percentile.

The false alarm probability (FAP) of the final orbital solution was derived by repeating the whole hybrid  analysis on $10^5$ sets of scrambled data.

\subsection{BD+49~828}

\begin{figure}
\centerline{\includegraphics[angle=0,scale=0.5]{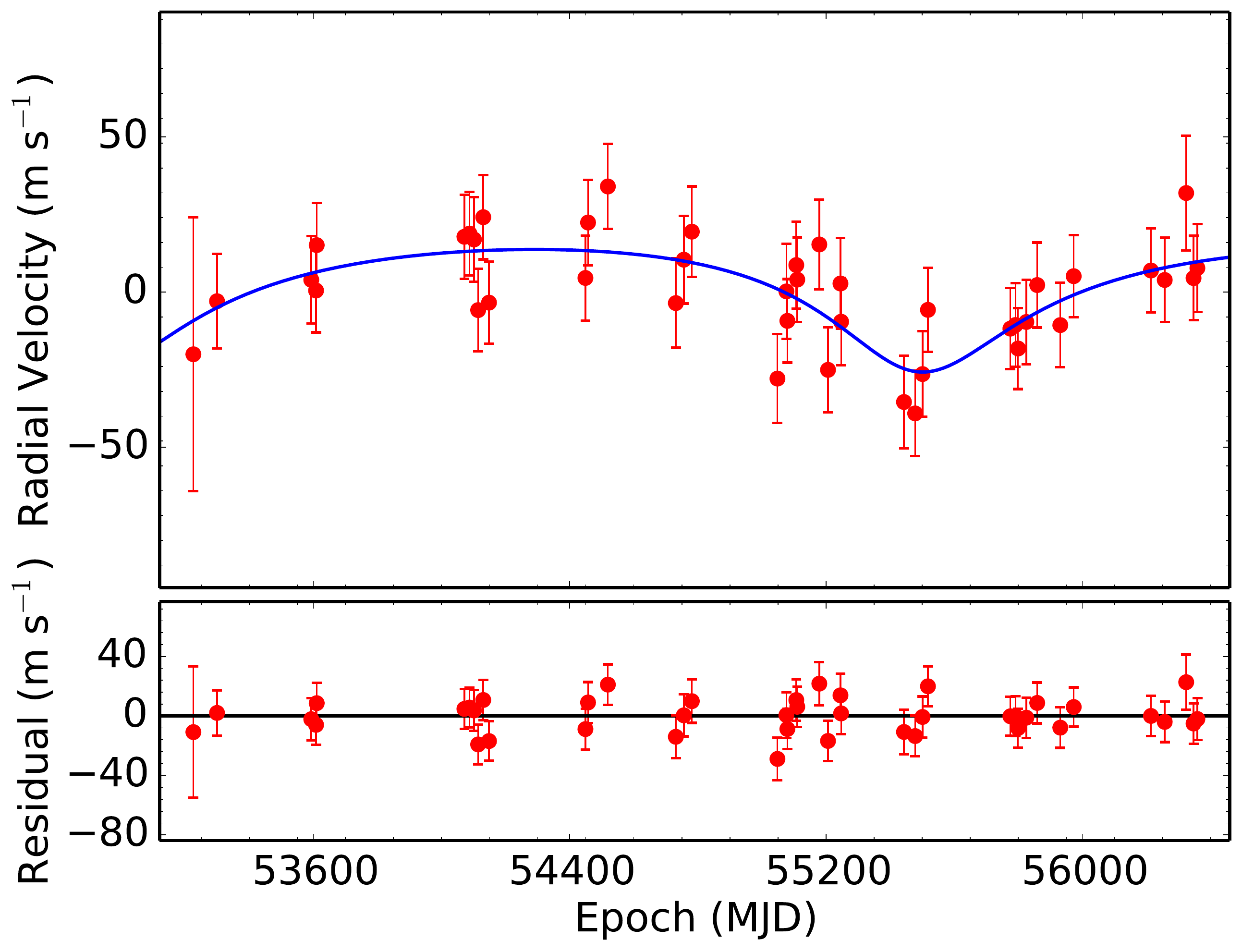}}
\caption{Radial velocities of BD+49~828 as in Table \ref{Tab:Table1} plotted as a function of time together with  the solution detailed in Table~\ref{Tab:Table5}.}
\label{figrv1}
\end{figure}

The  RV measurements for this star are presented in Table \ref{Tab:Table1} and Figure \ref{figrv1}. 
The RV variability is characterized by an amplitude of $73 \mps$ and  mean uncertainty of $8.5 \mps$. 
The LS periodogram of these data (Figure \ref{figls1}) reveals only one significant peak at a period 
of $\sim2410$ days at the false alarm probability FAP$_{LS}$ $<$ 0.01.

The final fit of a Keplerian orbit to the data 
resulted in the post-fit rms residua of $11.6 \mps$ and jitter of $4.4 \mps$ in very good agreement with our estimates. 
The fit produced the solution with $\chi^2=1.35$, and, after $10^5$  trials the $\mathrm{FAP}<10^{-5}$. 
The final best-fit Keplerian parameters are listed in Table \ref{Tab:Table5} and the model orbit is plotted together with the data in Figure \ref{figrvp1}. 
No additional periodic signal was detected in the post-fit RV residua (Figure \ref{figls1}).

This solution indicates the presence of a $m_{2} \sin i=1.6 ^{+0.4}_{-0.2} \Mjup$ body in a $4.2^{+0.32}_{-0.2}$~AU, 
moderately eccentric, $e=0.35^{+0.24}_{-0.10}$ orbit with the period of $P=2590 ^{+300}_{-180}$ days. 
Obviously, the long orbital period of BD+49~828 b is currently not very well constrained. 
Clearly, no less than 8 years of additional observations are needed to determine it more precisely.

\subsection{HD~95127}

\begin{figure}
\centerline{\includegraphics[angle=0,scale=0.5]{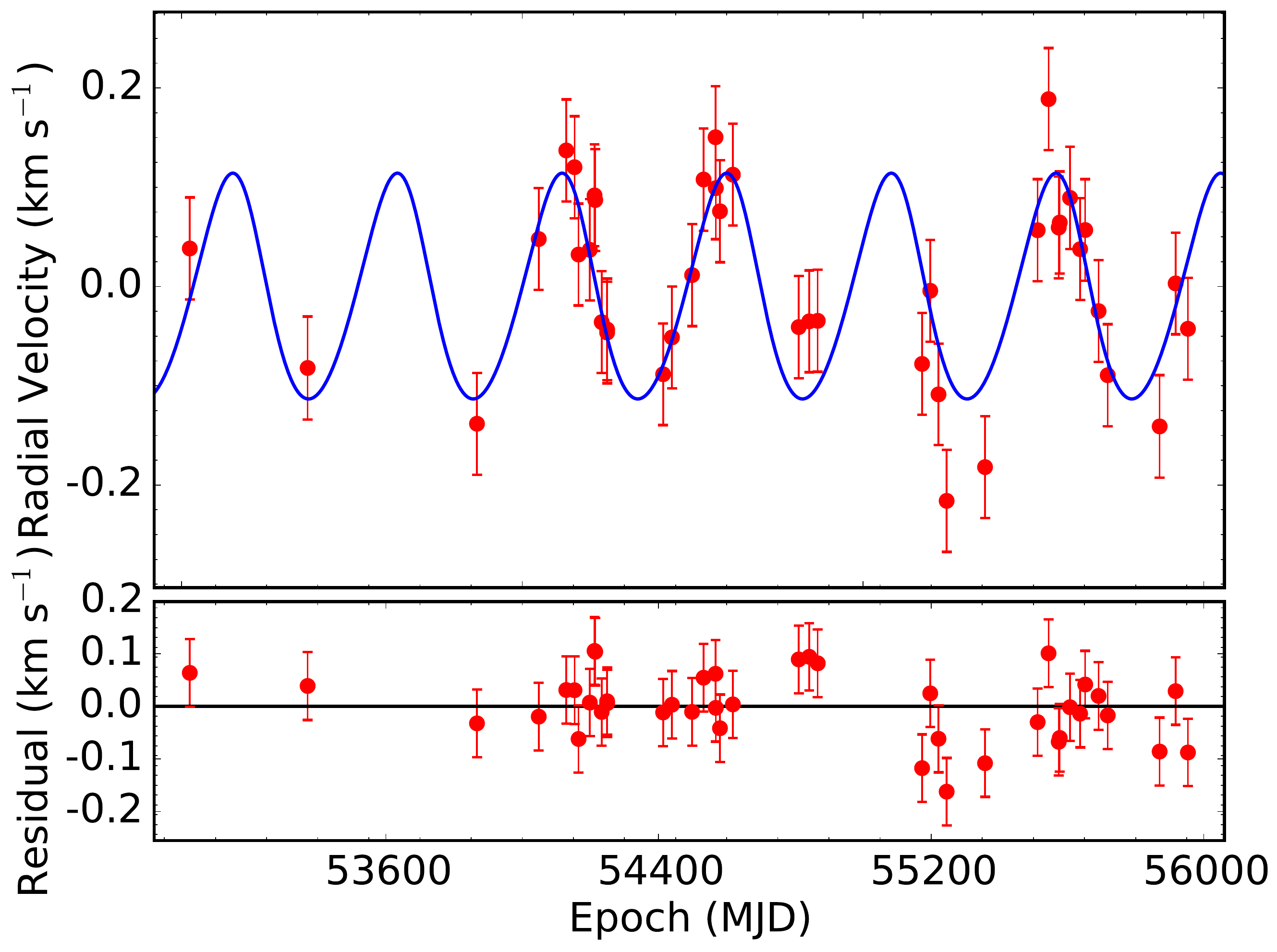}}
\caption{Radial velocities of HD~95127 as in Table \ref{Tab:Table3} plotted as a function of time together with  the solution detailed in Table~\ref{Tab:Table5}.}
\label{figrv2}
\end{figure}

RVs of HD~95127  (Table~\ref{tab-rv_bs_measurements-HD_95127} and Figure \ref{figrv2}), with an amplitude 
of $404.8 \mps$ and  estimated mean RV uncertainty 
of $5.98 \mps$,  
reveal a clear 483-day periodicity ($\mathrm{FAP}_{LS}<$0.001) in the corresponding LS periodogram  (Figure~\ref{figls2}). 

The orbital solution includes a planet with the minimum mass of $m_{2} \sin i$=$5.01^{+0.61}_{-0.44}\Mjup$, 
in a $482^{+5}_{-5}$ day, $1.28^{+0.01}_{-0.01}$~AU, $e = 0.11^{+0.15}_{-0.06}$ orbit around the star 
with the $\chi^2=1.14$ and FAP$<10^{-5}$ (Table \ref{Tab:Table5}, Figure \ref{figrvp2}).  
The post-fit residua show no significant periodicities in the corresponding LS periodogram (Figure~\ref{figls2}).

We also note that the $47.5\mps$ jitter  and $50.9 \mps$ post-fit residua from the final model are  consistent 
with the stellar jitter interpreted as under-sampled $p$-mode oscillations 
(highly skewed distribution of jitter in our bootstraping analysis prevented a reliable uncertainty estimate).

\subsection{HD 216536}

\begin{figure}
\centerline{\includegraphics[angle=0,scale=0.5]{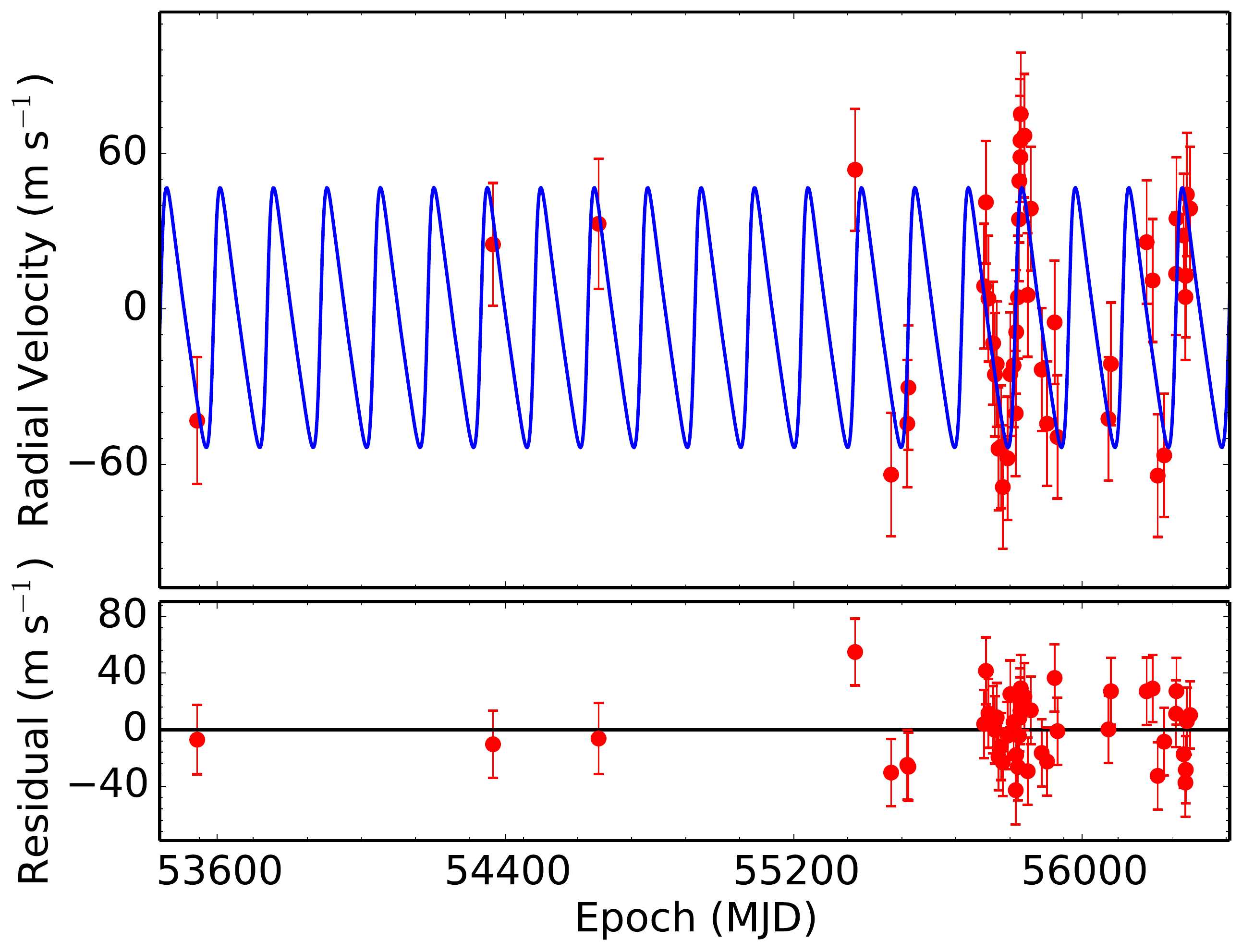}}
\caption{Radial velocities of HD~216536 as in Table \ref{Tab:Table3} plotted as a function of time together with  the solution detailed in Table~\ref{Tab:Table5}.}
\label{figrv3}
\end{figure}

\begin{figure}
\centerline{\includegraphics[angle=0,scale=0.5]{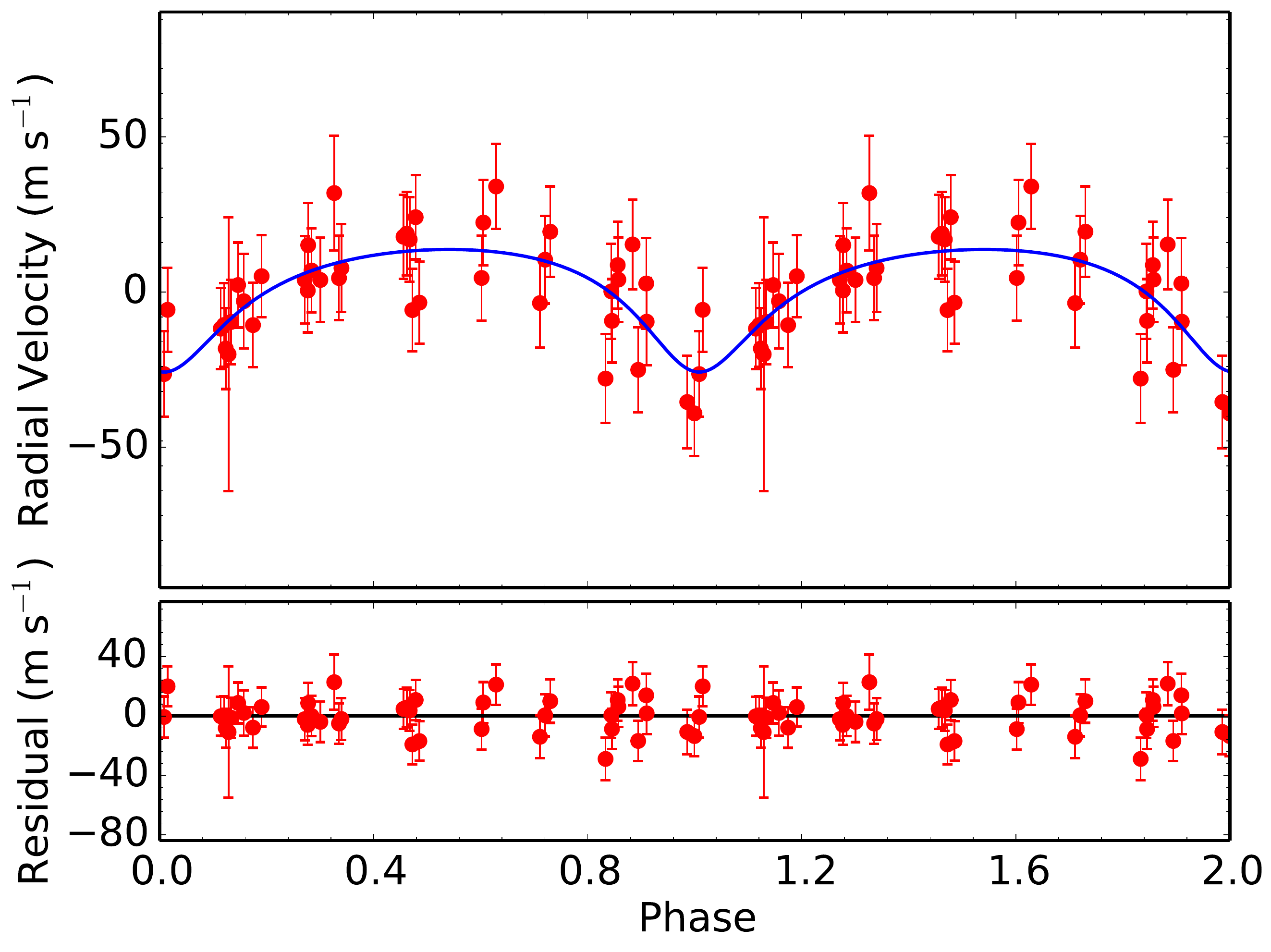}}
\caption{Radial velocities of BD+49~828 plotted as a function of orbital phase together with  the solution detailed in Table~\ref{Tab:Table5}.}
\label{figrvp1}
\end{figure}

\begin{figure}
\centerline{\includegraphics[angle=0,scale=0.5]{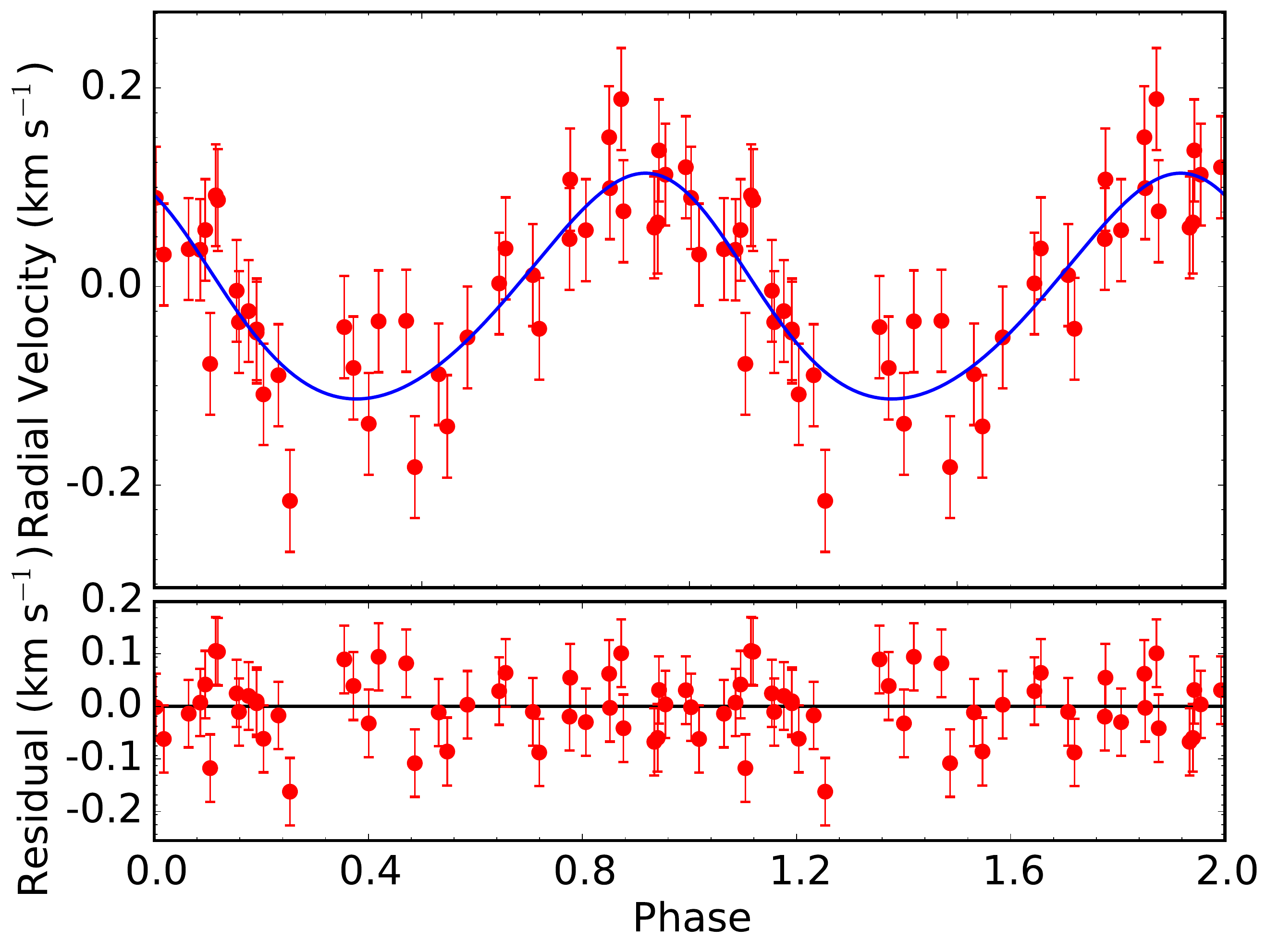}}
\caption{Radial velocities of HD~95127  plotted as a function of orbital phase  together with  the solution detailed in Table~\ref{Tab:Table5}.}
\label{figrvp2}
\end{figure}

The RVs for HD 216536 are presented in Table \ref{Tab:Table3} and plotted in Figure \ref{figrv3}. 
In this case, the mean uncertainty and amplitude characterizing the RV variability  are $6.4 \mps$ and $144 \mps$, respectively.
The LS periodogram of the data (Figure \ref{figls3}) shows a single significant period of 148 days with the  $\mathrm{FAP}_{LS}<$0.001.

\begin{figure}
\centerline{\includegraphics[angle=0,scale=0.5]{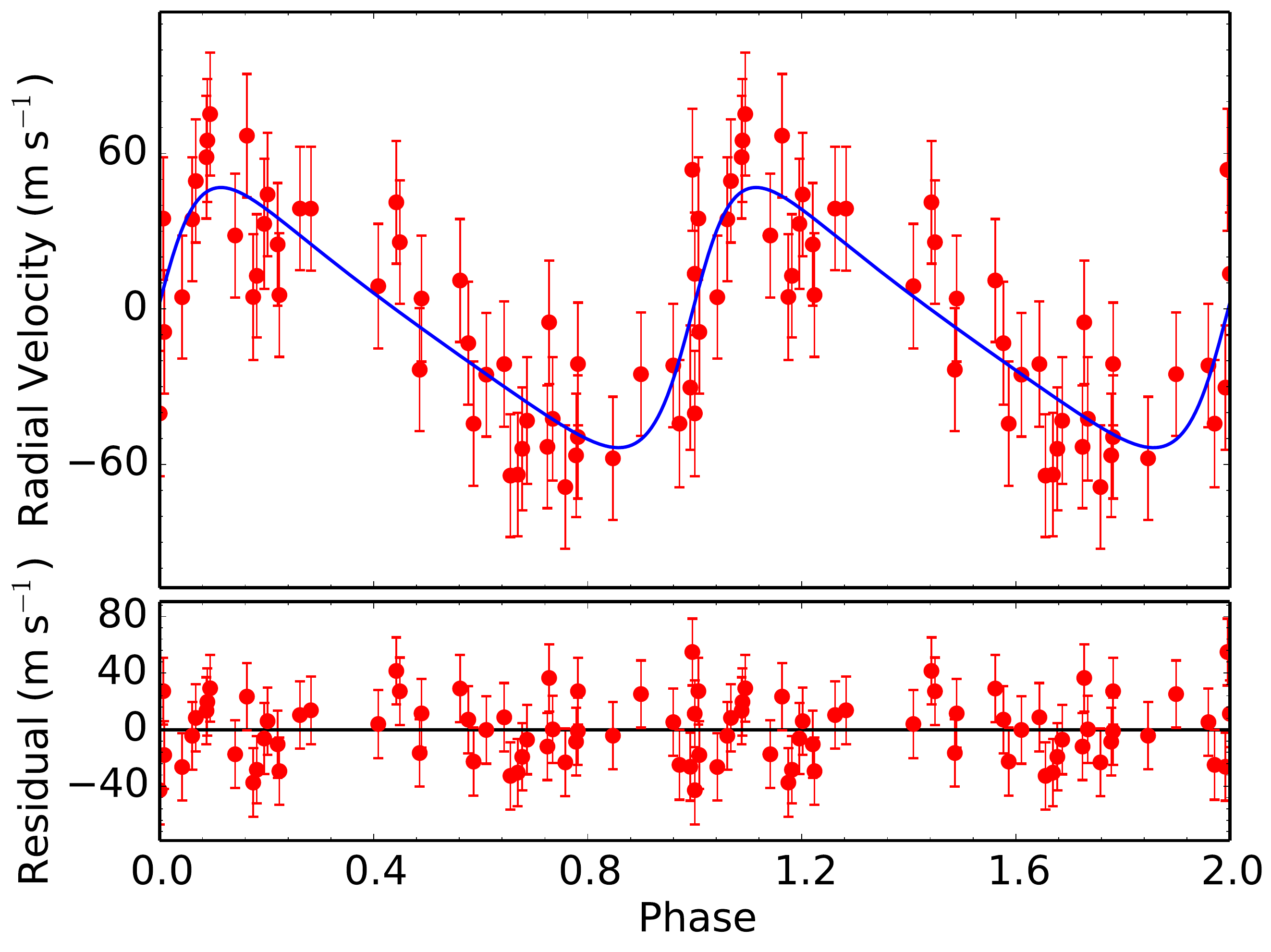}}
\caption{Radial velocities of HD~216536 plotted as a function of orbital phase  together with  the solution detailed in Table~\ref{Tab:Table5}.}
\label{figrvp3}
\end{figure}

The hybrid  modelling of these data, 
resulted in Keplerian  fit with the post-fit rms residua of $23 \mps$ and a jitter of $17.9 \mps$ (again presenting a highly skewed distribution), 
consistent with our estimates of the amplitude of unresolved $p$-mode oscillations based on stellar mass and luminosity. 
The final, best-fit model  ($\chi^{2}$= 1.29, FAP$<10^{-5}$) parameters are listed in Table \ref{Tab:Table5} and the model RV curve is plotted in Figure \ref{figrvp3}. 
No other periodic signal is present in the RV residua.

A minimum mass  of the companion to HD 216536 is $m_{2} \sin i=1.47^{+0.20}_{-0.12}\Mjup$, 
and its close-in, $0.609^{+0.002}_{-0.002}$~AU, $P=148.6^{+0.7}_{-0.7}$ day orbit has a moderate eccentricity, $e=0.38^{+0.12}_{-0.10}$.

\section{Stellar activity analysis: line bisectors, H$\alpha$ variations  and photometry.}\label{activity}

Both RV searches for planets and studies of star-planet interaction are affected by the intrinsic variability of host stars causing RV or BS variations not related to stellar reflex motion caused by the presence of an orbiting planet. Changes in spectral line shapes arising from motions in the stellar atmosphere, related to nonradial pulsations or inhomogeneous convection and/or spots combined with rotation can mimic low-level long period RV variations. 
Significant variability of red giants has been noted already by \cite{1954AnHar.113..189P} and \cite{1989ApJ...343L..21W}  
and made the nature of these variations a topic of numerous studies. 

\cite{1993ApJ...413..339H} showed that the low-amplitude, long-period RV variations may be attributed to pulsations, 
stellar activity - a spot rotating with a star, or low-mass companions while the presence of short-period, possibly $p$-mode oscillations 
as demonstrated in \cite{1994ApJ...432..763H}.  

The solar-type, $p$-mode oscillations are easily observable in high precision, photometric time-series measurements 
and they have been intensely studied based on the COROT  \citep{2006ESASP1306...33B} and KEPLER \citep{2010PASP..122..131G} data, 
leading up to precise stellar mass determinations \citep{2009Natur.459..398D, 2010ApJ...713L.176B, 2010A&A...522A...1K, 2011MNRAS.414.2594H}.
With the typical timescales for giants, ranging from hours to days, such oscillations typically remain unresolved in low-cadence 
observations focused on long-term RV variations and they contribute as an additional uncertainty to the RV measurements. 

Long-period RV variations in some giants have been successfully demonstrated to be due to the presence of low-mass companions
 \citep{1988ApJ...331..902C, 2003ApJ...599.1383H, 1993ApJ...413..339H, 2006A&A...457..335H, 2000ApJ...536..902V, 2001ApJ...555..410B, 2002ApJ...576..478F}. 
According to the exoplanets.eu website, 55 giant stars with substellar-mass companions have been identified so far. 
However, in some giants, the nature of the observed RV variations indicating the presence of long-term, secondary 
periods \citep{1954AnHar.113..189P, 1963AJ.....68..253H} remains a riddle \citep{2009MNRAS.399.2063N}.

Therefore, a reliable planet detection requires all the known sources of the observed RV variations to be investigated in a considerable detail. 
We note, however, that in the case of the three stars discussed in this paper, substantial eccentricities of the best-fit Keplerian orbits 
alone make stellar pulsations based alternative explanations less likely.

Ca~II H\&K resonance lines, if measurable, are good indicators of stellar activity. Unfortunately, in the HET/HRS spectra of red giants these lines are too weak to be usable. 
Therefore, we have developed a cross-correlation technique to monitor stellar activity with BS  measurements derived from exactly the same spectra 
that have been used for RV determinations \citep{Nowak2012, Nowak2013}. 

As an independent  stellar activity indicator we  also measured the H$\alpha$ index
($I_{\mathrm{H}\alpha}$) from HET/HRS red spectra based on the approach presented by
\cite{2012A&A...541A...9G} and \citet[][and references
therein]{2013ApJ...764....3R}, following the procedure described in detail in
\cite{2013AJ....146..147M}. 

Both activity indicators, BS and $I_{\mathrm{H}\alpha}$ were computed at the same epochs and from the same spectra as the RVs but such subtle effects 
as instrumental profile variations were not  accounted for in these measurements. Consequently,  they are, to some extend, subject to instrumental effects 
presumably not present in the RV measurements. As the $H\alpha$ line is also rather weak in K giants it is especially prone to  instrumental effects. 
To monitor such effects we also measured  the index
of the  Fe~I 654.6239 nm line
($I_{\mathrm{Fe}}$), which is insensitive to stellar activity. 

Also, the PTPS stars are relatively bright, which makes their photometric measurements available from the published surveys. 
These three kinds of data provide most of the information that we use to diagnose any possible activity of our target stars 
that may mimic RV variations due to the Keplerian motion.

\subsection{BD+49~828}

The maximum rotation period of BD+49~828, given the estimated projected rotation velocity of $\vrotsini = 1.7
\pm 0.8 \kmps$ and stellar radius of $7.6 \pm 1.3 \Rsun$,  is  $226\pm114$ days, much less than the orbital period, which makes rotation of a spot on the stellar surface an unlikely scenario. 

\begin{figure}
\centerline{\includegraphics[angle=0,scale=0.5]{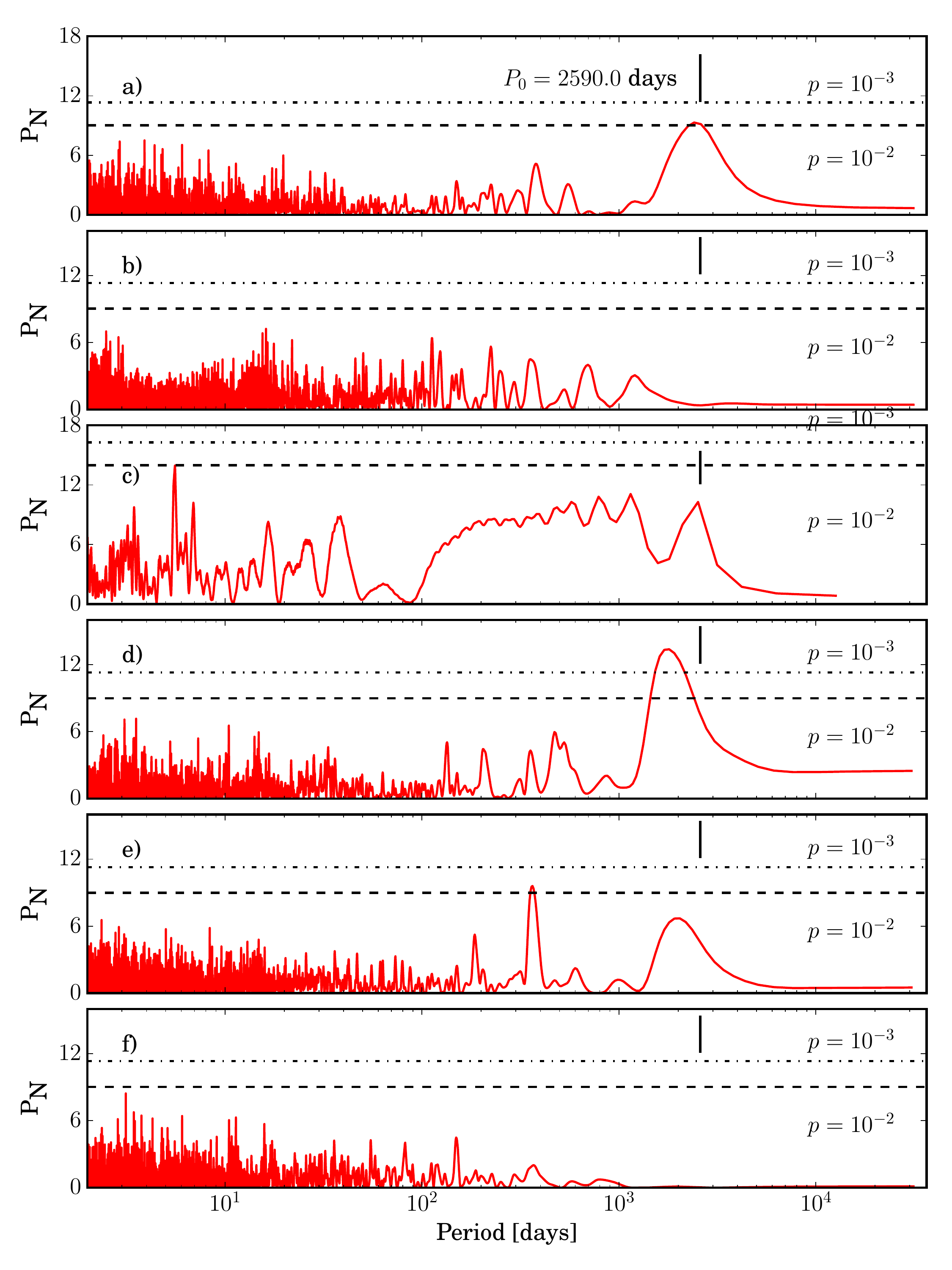}}
\caption{Lomb-Scargle periodograms for  BD+49~828. From top to bottom: HET RV data, BS, WASP photometry \citep{2010A&A...520L..10B}, $I_{\mathrm{H}\alpha}$, $I_{\mathrm{Fe}}$,  and RV post-fit residua. The low-signal peak in the periodogram for WASP photometry is most certainly an artifact as the available data cover only a fraction of the orbital period present in the RV data - see Sect. 4.1 for discussion.}
\label{figls1}
\end{figure}

There are two archival photometric 
time-series available for this star. The Northern Sky Variablity Survey   (NSVS; \citealt{2004AJ....127.2436W}) contains 
a set of 81 epochs of photometric measurements made over 216 days between MJD 51414 and 51630, which give the mean brightness 
and rms scatter of 9.23 mag and 0.028 mag, respectively. The Wide Angle Search for Planets (WASP; \citealt{2010A&A...520L..10B}) 
archive contains a much larger data set of 5557 epochs of photometric observations made over 1262 days between MJD 53196 and 54458. 
These measurements are partially contemporaneous with our RV measurements and are characterized by the respective 
mean brightness and the rms scatter of 9.659 mag and 0.019 mag. None of these data series contain a periodic signal 
close to the best-fit Keplerian orbital period. The LS periodogram of the WASP observations is shown in Figure \ref{figls1}. 
An absence of any periodic signal in the existing photometric time series suggests that pulsations or a spot rotating with the star cannot be the source 
of the observed RV variations. However, the time covered by photometric observations is short compared to orbital period. Therefore we conservatively assume that all the observed photometric scatter originates 
from a spot of a size of $f=1.9\%$ of the rotating stellar surface. Then, according to \cite{2002AN....323..392H}, we might 
expect a periodic variation of the RVs at a full amplitude of $64 \mps$.  This would be similar to the observed RV variations 
and it would generate BS variations at the amplitude of $45 \mps$ correlated with RV. To resolve such a situation one needs to carry out 
a detailed study of BSs and search their variations for periodicities that would correlate with those observed in the RV measurements.

The BS measurements for this star have a mean value of $18\pm33 \mps$ and  amplitude of $142 \mps$ with the mean 
uncertainty of $19 \mps$. The Parsons correlation coefficient between the RV and the BS variations amounts to -0.09 indicating that they are independent.  
In addition,  the BS variations do not exhibit any measurable periodicities (the LS periodogram of the BS variations 
is shown in Figure \ref{figls1}.) Finally, the $2590^{+300}_{-180}$-day orbital period of BD+49~282 b is much too long 
to be related to the observed rotational period of the star, which is more than 11 times shorter than that.

Our $I_{\mathrm{H}\alpha}$ measurements for BD+49~828 present a mean value of 0.0365 and range 
between 0.0381 and 0.0353, i.e. explicit variations of $7.5\%$. 
At the same time $I_{\mathrm{Fe}}$ has a mean value of 0.01147 and ranges from 0.01169 to 0.01132 i.e. varies by $3.3\%$ 
accounting for nearly a half of the observed $I_{\mathrm{H}\alpha}$ variations. 
The LS periodograms of both indices are shown in Fig. \ref{figls1}.  The $I_{\mathrm{H}\alpha}$ shows a significant periodicity of about 1750 days ($\sim 5$~yr)  
and $I_{\mathrm{Fe}}$ data show a signficant period of about 1 yr, apparently of seasonal origin. The long period  signal 
is well separated from the RV period of $2590^{+300}_{-180}$~days. The 1~yr signal is, however, close to the estimated rotation period of the star.  
Both LS of line profile indices show a similar pattern suggesting strong influence of instrumental effects enhanced 
in the case of $I_{\mathrm{H}\alpha}$ by stronger signal-to-noise ratio. Indeed, the Parson's correlation coefficient 
between RV and $I_{\mathrm{H}\alpha}$ is r=-0.344 only while the correlation between $I_{\mathrm{H}\alpha}$ and $I_{\mathrm{Fe}}$ 
is r=0.620, much above the critical value of 0.408 (confidence level of 0.01). We may safely assume  therefore that 
the observed $I_{\mathrm{H}\alpha}$ variability is induced by instrumental effects and carries no evidence of stellar activity.

Based on the discussion above, we conclude that the RV variations in BD+49~828 cannot be explained 
in terms of rotation induced stellar activity or pulsations. Consequently, a Keplerian modeling of the measured 
RV variations of the star is justified and the presence of an orbiting low-mass companion represents 
the most natural interpretation of the data. The observed BS  variations must originate from another 
source which is either random or not resolved with our observations. 
The origin of $I_{\mathrm{H}\alpha}$ and $I_{\mathrm{Fe}}$ variations seems to be instrumental but it's nature is not clear.

\subsection{HD~95127}

With the estimated projected rotation velocity of $\vrotsini = 2.6 \pm 0.7 \kmps$ and radius 
of $20 \pm 9 \Rsun$ the maximum rotation period of HD~95127 is $389\pm204$~days, much less that the orbital period.

\begin{figure}
\centerline{\includegraphics[angle=0,scale=0.5]{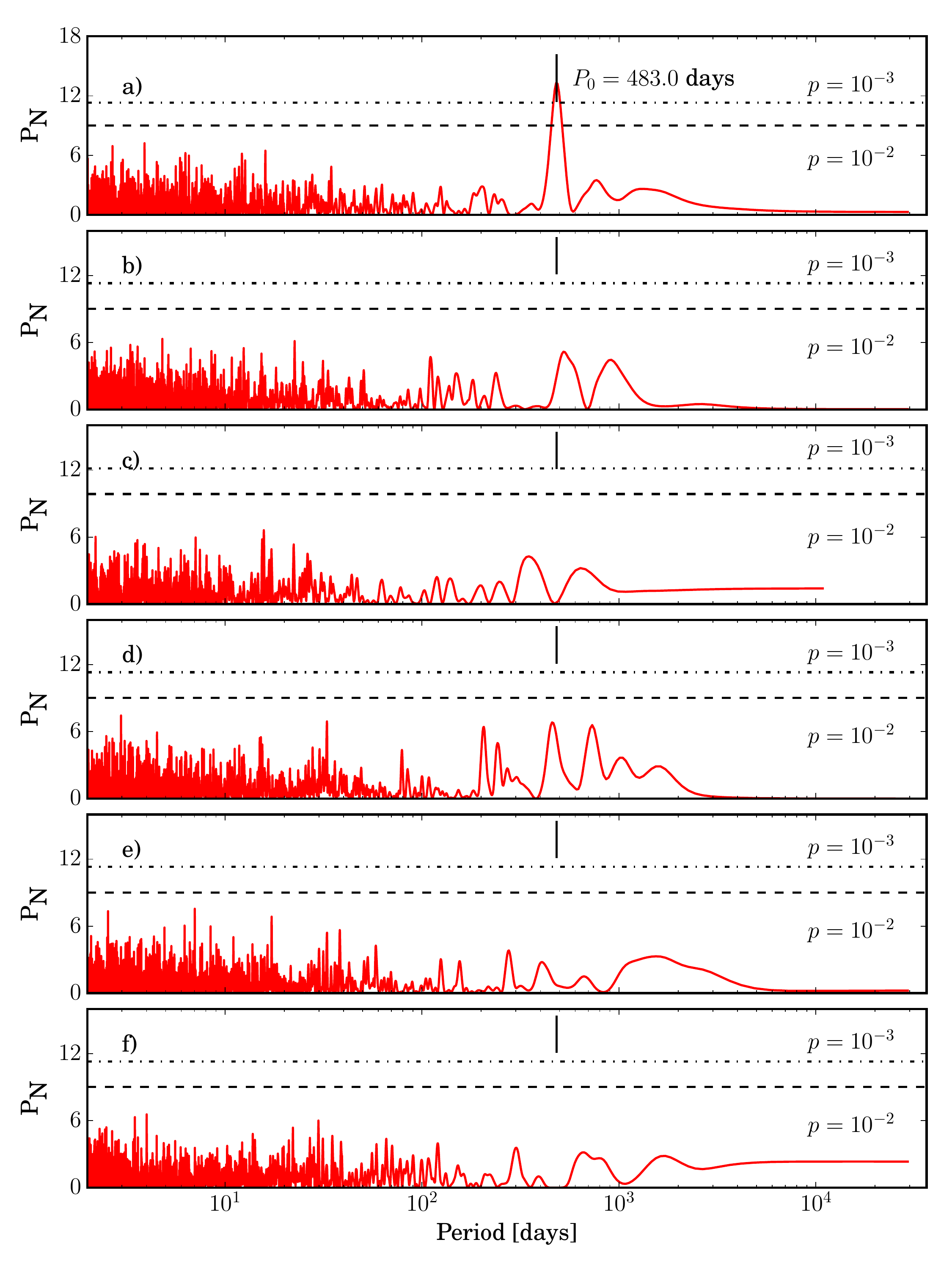}}
\caption{Lomb-Scargle periodograms for   HD~95127. From top to bottom: HET RV data, BS, Hipparcos  photometry \citep{1997ESASP1200.....E}, $I_{\mathrm{H}\alpha}$, $I_{\mathrm{Fe}}$,  and RV post-fit residua.}
\label{figls2}
\end{figure}

\begin{figure}
\centerline{\includegraphics[angle=0,scale=0.5]{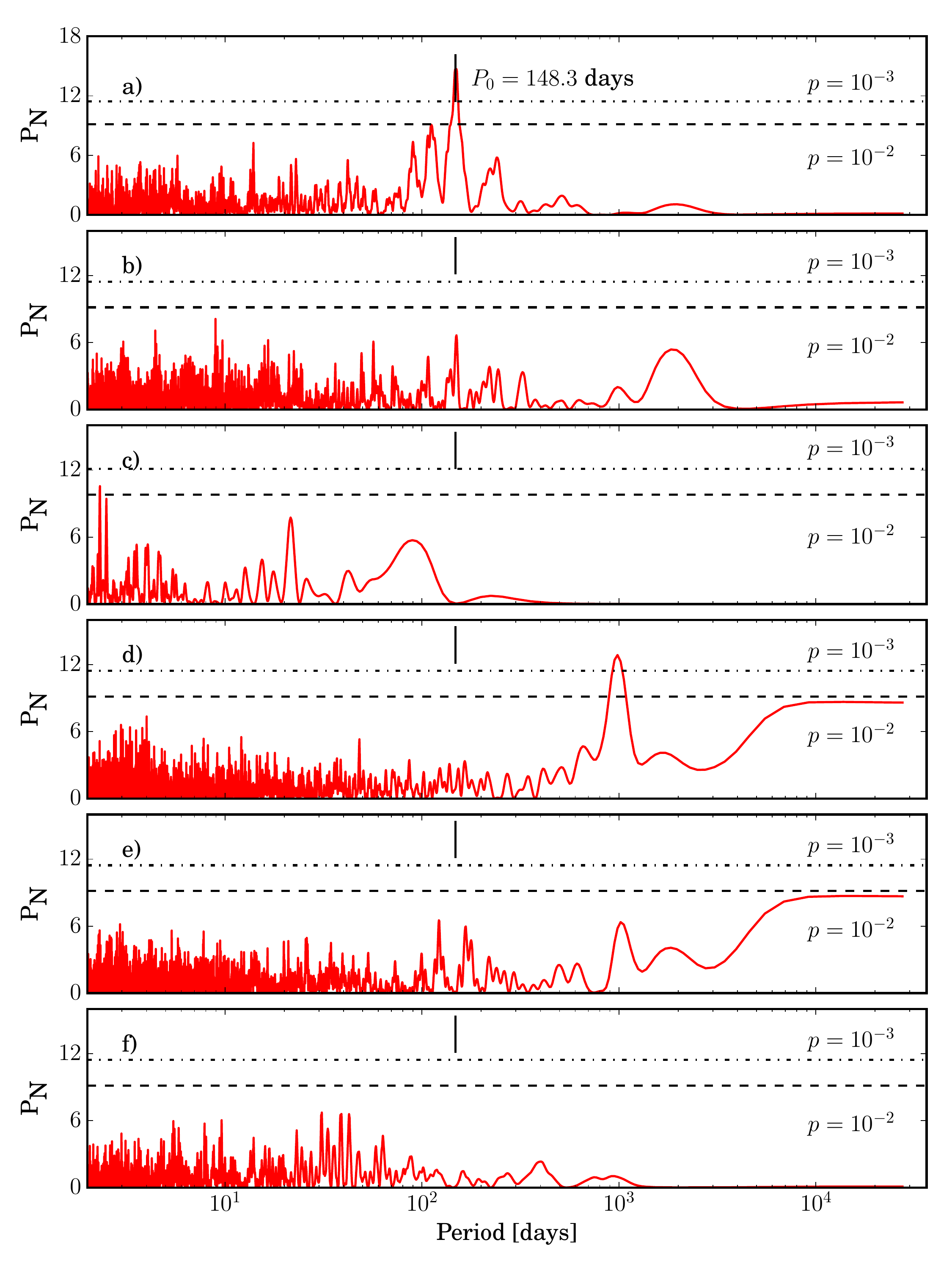}}
\caption{Lomb-Scargle periodograms for  HD~216536. From top to bottom: HET RV data, BS, NSVS photometry \citep{2004AJ....127.2436W}, $I_{\mathrm{H}\alpha}$, $I_{\mathrm{Fe}}$,  and RV post-fit residua.}
\label{figls3}
\end{figure}

If the observed RV variations were indeed a result of a spot rotating with the star the observer RV semi-amplitude would 
require a spot of $9\%$ of the stellar surface \citep{2002AN....323..392H}. Such a spot would be easily detectable 
with moderate quality photometry. The only existing extensive $V$-band photometric monitoring for HD~95127 comes 
from  SWASP \citep{2006PASP..118.1407P} where 540 epochs are available over only about 100 days between 
MJD 54091.1 and 54194.9. The mean brightness of the star was $8.394\pm0.012$~mag and the data show 
an amplitude of 0.033~mag only, three times less than expected. 95 Hipparcos \citep{1997ESASP1200.....E} 
$H_{\mathrm{p}}$ measurements taken over three years between MJD 47877 and 48961 show also a constant 
brightness of $H_{\mathrm{p}}=8.307\pm0.014$~mag and no significant periodicity (Fig. \ref{figls2}). 
We can conclude therefore that the existing photometric data do not support the spot hypothesis.
The lack of any significant periodic signals in photometric data excludes also pulsations as a possible source of RV variations.

The spot hypothesis predicts BS variations of semi-amplitude of $23.7\mps$ \citep{2002AN....323..392H} while 
the observed BS semi-amplitude is $42\mps$. However, the uncertainties in BS are much larger than 
in RV ($16.6\mps$ on average) and the mean value of BS is $23\pm23\mps$. 
The LS analysis of BS data shows no trace of periodicity at the rotation period or at  the Keplerian best fit period (Fig. \ref{figls2}).
Moreover, the Parson's correlation coefficient between RV and BS is r=0.25 with the critical value of 0.4, 
and no relation between these quantities can be stated. Again, the BS data do not support the spot hypothesis.

The $I_{\mathrm{H}\alpha}$ index for HD~95127 has a mean value of 0.0339, and  varies by 8.5$\%$. At the same time $I_{\mathrm{Fe}}$ index presents a mean value of 0.00983 
 and varies only by 1.9$\%$. None of indices shows any significant periodic signal (Fig. \ref{figls2}).
Also the correlation analysis carries no useful information on activity as both correlations between 
RV and $I_{\mathrm{H}\alpha}$ and between $I_{\mathrm{H}\alpha}$  and $I_{\mathrm{Fe}}$, r=0.419 and r=0.424, 
respectively are at the level of critical value (0.403).

Therefore we conclude  that the only significant periodic signal  present in the RV  data is due to an orbiting planet and there is no indication 
of pulsations or spot rotating with the star that can be inferred from our data.

\subsection{HD 216536}

The estimated projected rotation velocity of $\vrotsini = 2.6 \pm 0.5 \kmps$ and the radius of $12.5 \pm 4.0 \Rsun$ 
of this star result in the maximum rotation period of  $243\pm91$~days. HD 216536 had been observed by 
the NSVS \citep{2004AJ....127.2436W} 5.6~years before the beginning of our measurements.
Photometric observations of the star at 90 epochs between MJD 51304 and 51483 have been collected.
These measurements are characterized by the mean brightness of 9.118~mag and the rms scatter of 0.029~mag.  
As shown in the LS periodogram of Figure \ref{figls3}, no significant periodicity is present in the data.
The observed photometric scatter could be explained in terms of a f$=2.9\%$ spot rotating with the star. 
This would result in periodic, $185 \mps$ and $25 \mps$ variations in the RVs and BSs, respectively \citep{2002AN....323..392H}. 
The fact that the predicted, spot generated RV variations are similar in amplitude 
to those actually observed requires a careful analysis of the BS  behavior.

The measured BSs exhibit variations that have the mean value of $14\pm 22 \mps$,  the amplitude of $106 \mps$, 
and the mean uncertainty of $14 \mps$. The Parsons correlation coefficient of the RV and BS variations amounts
 to 0.29 which means that they are not likely to be correlated.

In addition, BS variations do not show any measurable periodicities. Consequently, the absence of any correlation 
between RVs and BSs and lack of periodic variations in both BS and existing photometry invalidates the spot hypothesis.

Our measurements of $I_{\mathrm{H}\alpha}$ index for HD 216536 reveal a mean value of 0.0354, 
and amplitude of variations of only 0.0021 ($6\%$). 
The $I_{\mathrm{Fe}}$ variations are even weaker - 2.6\%.  Both  indices show very similar pattern 
of LS and there is a significant signal at 984 days in $I_{\mathrm{H}\alpha}$  data. The signal is well separated 
from the Keplerian period and the estimated maximum rotation period of 347$\pm$115 days. The Parson's correlation
 between $I_{\mathrm{H}\alpha}$ and $I_{\mathrm{Fe}}$, r=0.595 is above critical (r=0.376) 
 which proves the instrumental nature of the signal. The RV variations are unrelated 
 to that effect as the correlation coefficient between RV and $I_{\mathrm{H}\alpha}$ is only r=0.055.

We conclude that, according to the available evidence, the RV variations in HD~216536 cannot be related to rotation induced 
stellar activity or stellar pulsations. Consequently, it is the most logical to assume that these variations are due to 
the orbital motion of a planet-mass companion around the star. As for the other two stars, the observed BS variations 
must have another, unrelated source which is either random or not resolved in time with our observations.

\section{Discussion. Planetary systems of evolved stars.}\label{conclusions}

The three new stars with planets presented here, provide a representative sample of the wealth of 
planetary system architectures observed in the evolved systems and illustrate how searches for planets 
around such stars can address various aspects of stellar and planetary systems evolution. 

By exploiting the approach of \cite{2007ASPC..371..189F} and \cite{ 2011ApJS..197...26J} in Keplerian orbit fitting 
we obtained  independent estimates of stellar jitter, whose amount is consistent with the  amplitudes of $p$-mode 
oscillations calculated from the scaling relations of \cite{KjeldsenBedding1995}. 
This proves that the L/M ratios for the target stars  and hence their positions on the HRD are correct.

HD~95127, a bright, $\llsun=2.28 \pm 0.38$, solar-mass $M = 1.2 \pm 0.22 \Msun$ giant, with  
$R = 20 \pm 9 \Rsun$ and  [Fe/H]$=-0.18\pm0.05$  
hosts a $m\sin i = 5.01^{+0.61}_{-0.44}\Mjup$ planet in a $1.28^{+0.01}_{-0.01}$~AU, $e=0.11^{+0.15}_{-0.06}$  orbit.
The relatively low mass of the star and its expected large radius at the tip of the Red Giant Branch (RGB) set the current orbit 
of its planetary companion well within the engulfment zone 
\citep{2009ApJ...705L..81V, Kunitomo2011, Villaver2014}. This  
 presents an independent proof that the star is still evolving up the RGB.

HD 216536, a $M / \Msun = 1.36 \pm 0.38$ and [Fe/H]$=-0.17\pm0.09$ star, within the parameter 
 uncertainties, is also evolving up the RGB. Most likely  this star is still before the helium ignition.
It harbors one of the most close-in planets orbiting a giant, which supports our conclusion concerning its evolutionary stage.
Given its $0.609^{+0.002}_{-0.002}$~AU orbit, this $1.47^{+0.20}_{-0.12}\Mjup$ planet  is one 
of the very few warm Jupiters around giants, prone to strong  tidal interactions 
and most likely to be ingested  before its host reaches the tip of the RGB
\citep{2009ApJ...705L..81V, Kunitomo2011, Villaver2014}. 

The position of BD+49~828 on the HRD suggests that this $\mathrm{M} / \Msun = 1.52 \pm 0.22$, slightly metal deficient
([Fe/H]=-0.19$\pm$0.06) star, is evolving up the RGB,  possibly undergoing the first dredge-up. 
Its $m \sin i=1.6^{+0.4}_{-0.2}\Mjup$ companion, $4.2^{+0.32}_{-0.2}$ AU away of the star, 
is one of the most distant planets orbiting giants detected so far.
It is very likely that the  planet will not be affected by stellar evolution 
and similar to HD 4732 c  \citep{2013ApJ...762....9S} or HD 120084 b  \citep{2013PASJ...65...85S}. 
Taking into account the stars's uncertain luminosity estimate, the planet may also 
be located within the habitable zone as defined in  \cite{2013ApJ...765..131K}. 
 
\begin{figure}
\centerline{\includegraphics[angle=0,scale=.75]{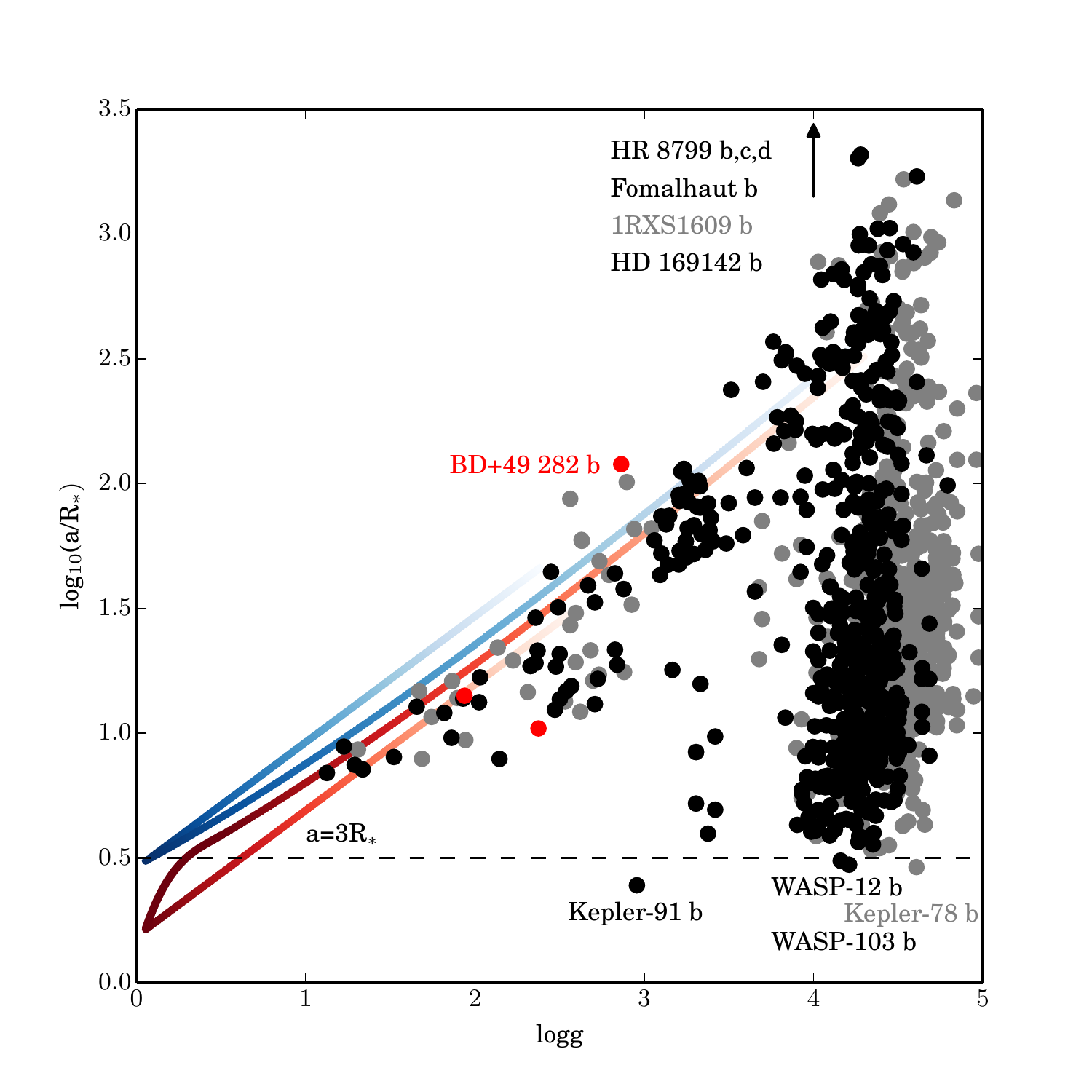}}
\caption{Orbital separation - log(a/R$_{\star}$) for planets orbiting hosts at various evolutionary stages - log(g), all data from exoplanets.eu.  Grey symbols - all planets, black symbols - planets around $1-2 \Msun$ stars, red symbols - new planets presented here. Red and blue lines - minimum distance to avoid engulfment for a $1.5 \Msun$ star and $1 \Mjup$ planet  system and minimum orbit above which a planet is not sensitive to tidal interaction with the stellar host from \cite{Villaver2014}. See text for discussion.   }
\label{Discussion}
\end{figure}

To put the three planetary candidates discussed in a wider  perspective, we plotted them as red circles 
in the  log(g) versus  $\log(a/R_{\star}$) plane in Figure \ref{Discussion}. 
Because the orbital radius, $a/R_{\star}$, is a major factor scaling the strength of tidal star-planet interactions, this plot may be regarded as an illustration of the tidal evolution of planetary systems.
In the same Figure we have also placed all known exoplanets\footnote{All data from exoplanets.eu}  
and KOIs that orbit stars with masses in the range of 1-2~$\Msun$ as black circles.
Finally   the total sample of planetary companions, for which  
with at least  semi-major axes, and stellar masses and radii are available  is shown as grey circles. 
For all stars, including the three ones presented here, uniform $\log g$'s were calculated from stellar masses 
and radii adopted in the discovery papers. We note, that for some planetary systems hosts the spectroscopic 
$\log g$ do not match those calculated from stellar mass and radius. In particular this concerns 7~CMa \citep{2011ApJ...743..184W}, 
another long-period planet for which the host star $\log g_{M,R}=3.9$.

Also included in this Figure is the minimum distance to avoid engulfment for a $1 \Mjup$ planet orbiting a $1.5 \Msun$ star 
under the \cite{2005ApJ...630L..73S} mass-loss prescription (red line) 
and the minimal orbit beyond which the planet is not affected by the tidal forces  (blue line)
from \cite{Villaver2014}.  The dashed line shows the $a/R_{\star}=3$ border inside 
of which  tidal interactions are expected to lead  quickly to ingestion.
Note that the minimum distance to avoid engulfment for a $1 \Mjup$ planet orbiting a $1.5 \Msun$ star 
 from \cite{Villaver2014} (red line) fits the general slope of decreasing $a/R$ with $\log g$ 
 for known planetary systems in Figure \ref{Discussion}. In fact, up to a very late stellar evolution 
 phase at RGB where $\log g<1$ and tidal ingestion is happening this relation simply reflects the steady growth of a stellar radius.

An interesting feature of Figure \ref{Discussion} is that after the discoveries of 
HD~102956~b \citep{2010ApJ...721L.153J}, 
Kepler-56 b,~c  \citep{2013MNRAS.428.1077S}, 
Kepler-391~b,~c  \citep{2014ApJ...784...45R}, 
KOI-1299~b \citep{2014arXiv1410.2999C, 2014arXiv1410.3000O}  and
HIP~67851~b \citep{2014arXiv1409.7429J}, 
the gap in orbital separations of planetary systems orbiting subgiants (here: $\log g =3.5 \pm0.5$) 
pointed out in \cite{Nowak2013} no longer exists. A relatively low number of systems around that kind 
of stars is obviously a consequence of a low population of that part of the HRD (see for example Figure 3.5.5 in \citealt{1997ESASP1200.....E}).

The ingestion itself is supposed to be a quick process as clearly illustrated in Figure  \ref{Discussion}. Only three dwarfs (here: $\log g =4.5 \pm 0.5$) 
(Wasp-12~b \citep{2009ApJ...693.1920H}, 
 Kepler-78~b \citep{2013ApJ...774...54S},
 Wasp-103~b \citep{2014A&A...562L...3G}) 
out of 1037 ones or $\approx0.3\%$  are located in that range.  The same is true for only  one (Kepler-91~b)  out of 44 
or $\approx2\%$ giants (here: $\log g =2.5 \pm 0.5$).

The most evolved planetary systems with $\log g <1.5$, very close to engulfment, are:
HD~220074~b, HD~208527~b \citep{2013A&A...549A...2L},
$\beta$~Cnc~b \citep{2014A&A...566A..67L},
HD~96127~b \citep{2012ApJ...745...28G},
$\beta$~Umi~b \citep{2014A&A...566A..67L}, all within $a/R_{\star}=9$.
Interestingly enough, except for Kepler-91~b there is still no planet orbiting an evolved star with $\log g < 3$ 
within $a/R_{\star}< \approx 7$ (the  tightest orbit being HD 220074 b  at $a/R_{\star}=6.93$). This may suggest 
that tidal interactions between most extended evolved stars and their planets are more far-reaching.

It is important to note, that the part of Figure \ref{Discussion} occupied by bright giants, (here: $\log g =1.5 \pm 0.5$)
may be populated by both planets hosted by RG stars before RGB tip and engulfment, and post RGB-tip 
stars planets.
Although the latter planets have avoided tidal ingestion at the RGB tip, their orbits must have decayed  (channel (2) in \citealt{2009ApJ...705L..81V}, see discussion of their Figure 1), in some cases to well within 1 AU,
and may subsequently be subjected to ingestion during their  hosts Asymptotic Giant Branch evolution. 
Undoubtedly, all these stars deserve  
more detailed spectroscopic analysis to asses their evolutionary stages and hence the dynamical history of their planets in more detail.

All planets below the red line with at least one Jupiter mass are expected to be eventually ingested by their hosts 
before they reach the tip of the RGB. An obvious conotation from Figure \ref{Discussion} is that vast majority of known 
planetary systems is located within the minimum distance to avoid engulfment (red line) and will not survive 
RGB evolution of their hosts, in agreement with \cite{2009ApJ...700..832C} and \cite{2010MNRAS.408..631N}. 
Only very few known planetary systems contain planets in  orbits, in which are not affected 
by the tidal forces (above the blue line). This is  105 out of 1205 planets or $\approx9\%$  of planets in the {\it{exoplanet.eu}} catalog, 
for which at least  semi-major axes, $M_{\star}$ and $R_{\star}$ are available. 
Most of the safe planets orbit dwarfs (77 out of 1037 or $\approx7 \%$). A few such systems exist around subgiants 
(10 out of 99 or $\approx10 \%$) with the most evolutionary advanced of them being HR 228 b and c \citep{2013ApJ...762....9S}.
Only 5 (out of 44 or $\approx 11\%$) of such systems are present around giants: 
HD~139357~b \citep{2009A&A...499..935D}; 
$\nu$~Oph~c and omi~UMa~b  \citep{2012PASJ...64..135S};
HD~120084~b \citep{2013PASJ...65...85S};
HD~14067~b \citep{2014arXiv1409.6081W}.
 BD+49 828~b presented here belongs to that group as well. 
 
The obvious shortage of known safe planetary  systems is most likely a consequence of current observational limitations.
It  illustrates the importance of long-term planet search projects like, for example, PTPS, the Okayama Planet Search \citep{2005PASJ...57...97S}
or new projects like Friends of Hot Jupiters \citep{2014ApJ...785..126K}. It also has serious consequences
for  identification of asymmetric planetary nebular formation processes  \citep{1994MNRAS.270..734H, 1996ApJ...468..774S, 2002ApJ...571L.161L}.

The giant planets are formed outside the snow-line, at a distance of  6 - 12~AU in the case of $1-2~\Msun$ stars  \citep{2008ApJ...673..502K}.
They migrate inward \citep{2007ApJ...665.1381A} in the protoplanetary disks through type II migration but other migration scenarios:
planet-planet scattering \citep{1996Sci...274..954R, 2001Icar..150..303F} 
or Kozai migration \citep{1962AJ.....67..591K, 2003ApJ...589..605W} 
are also possible. Such planets  may actually become dominant when the  protoplanetary disk is dispersed. 
The efficiency of the migration process, and especially a relative efficiency of various mechanisms is basically unknown 
due to the lack of known long-period planets.  We note the importance of Rositer-McLaughlin effect studies in that context \citep{2010MNRAS.405.1867S}.

An inventory of planetary system at various evolutionary stages of their hosts
is presented in Table \ref{Tab:Table6}, in which four sets of data are shown: all stars for which all parameters 
listed in Table  \ref{Tab:Table6} are available, and a subset of these stars limited to $1-2 \Msun$, as well as subsets 
of those two sets with the RV detected planets only.  In case of dwarfs, we see the impact of short-period 
transit planets as a huge variation between the mean value and the median of minimum planetary 
semi-major axis, hence the need for a ,,RV limited'' subsamples. 

Several interesting features of the known sample of exoplanets emerge.
In  RV limited subsamples not much variation in semi-major amplitude or eccentricity is seen between 
planetary systems at various evolutionary stages of their hosts. Both a and e vary by about 
$1\sigma$ between dwarfs and bright giants ruling out tidal circularization during the stellar evolution.

Another important general feature of the known planet sample is the steady decrease of metallicity as we move 
from dwarfs to several billion years older giants. That reflects the changes in stellar populations we are dealing with,  
varying star formation rate  \citep{2014ApJ...791...92T} possibly reflected in the planet formation rate 
\citep{2002ApJ...567L.149B, 2012ApJ...751...81J, 2013MNRAS.431..972J,2014ApJ...788...62H}. 

As we move from dwarfs to more evolved hosts we deal with more massive stars on average as well.  
An average dwarf with a planetary system is a $\approx1\Msun$ (F or G spectral type) star while  an average 
subgiant is already a $\approx1.5\Msun$ (MS spectral type A-F) star, and a giant  (MS A-type star) may be almost twice 
as massive as a dwarf  (5-10$\sigma$ difference).
Stellar mass is not expected to increase during  MS and RSG evolution and it is clear that we must be facing
  a strong selection effect here. Even in  a narrow $1-2\Msun$   mass range we still see more massive, evolved stars.
On one hand, masses of single evolved stars are not easy to estimate, they may be simply overestimated \citep{2013A&A...557A..70M}, 
which would be the easiest explanation of the suspicious lack of solar-mass evolved planetary systems hosts.
On the other hand, taking into account that a  solar mass star is expected to lose $\sim 0.2-0.4 \Msun$ in the RG stage 
\citep{1966ApJ...144..108C, 1993AJ....105.1145F, 2005A&A...441.1117L, 2007ApJ...671..748K, 2011MNRAS.416L...6M} 
and remembering  that the fraction of RGB mass loss increases with stellar mass \citep{2008ApJ...676..594K},
we see that the known evolved planetary system hosts  
 might have been originally $\approx2-3\Msun$  MS stars, 
at the low -- intermediate  mass border or above. This points to their  very uncertain, possibly quite complicated  evolutionary history.

The planetary mass increase for more evolved hosts seems to be the most prominent feature present  in Table  \ref{Tab:Table6}.
An average giant hosts a companion about twice as massive than in the case of a dwarf, and a bright giant's companion is 3 times more massive, on average. 
Certainly, that is not an observational bias caused by RV jitter present in giants as companions below Jupiter mass 
can be easily detected around those stars (for instance BD+48~738~b, \citealt{2012ApJ...745...28G}).
Assuming an average value of $\sin i$,  companions  to bright giants are, on average, at the brown dwarf -- planet borderline (see also \citealt{2013A&A...555A..87M}).
It is rather obvious that such an amount of mass cannot be accreted by a giant planet during its host's RGB evolution \citep{1998Icar..134..303D}.

We note, however a lack of correlation between {\em{current}} host mass and its {\em{observed}} planetary system mass (Table  \ref{Tab:Table7}).
It does not necessarily mean that there is no relation between hosts and their planetary system masses postulated in \cite{2007A&A...472..657L}.  
We may speculate that, to show such a relation, one should use initial masses of the planetary system hosts, instead of the current ones, 
corrected for mass-loss over their entire evolution, and account for the planetary system mass - loss through planet ingestion or ejection. 
If we assume, for simplicity, that an average subgiant from Table \ref{Tab:Table6} (RV-limited, complete sample mean values of mass) 
lost $5\%$ of its initial mass, whereas a giant and a bright giant would lose  $15\%$ and  $35\%$, respectively, we obtain a perfect host mass -- planetary mass 
relation with a correlation coefficient  r=0.95, which is statistically significant at $99\%$ confidence level.  This  result is promising enough to justify a more detailed study.

\acknowledgments
We thank the HET resident astronomers and telescope operators for continuous 
support. 

AN, MoA,  BD-S, MiA and KK are currently supported by NCN grant 2012/07/B/ST9/04415.
MoA also acknowledges the ''Mobility+ III'' fellowship from the Polish Ministry of Science and Higher Education.
AW was supported by the NASA grant NNX09AB36G. 
GM acknowledges the financial support from the Polish Ministry of Science and Higher Education through the Iuventus Plus grant IP2011 031971.

The HET is a joint project of the
University of Texas at Austin, the Pennsylvania State University,
Stanford University, Ludwig-Maximilians-Universit\"at M\"unchen, and
Georg-August-Universit\"at G\"ottingen. 

The HET is named in honor of
its principal benefactors, William P. Hobby and Robert E. Eberly. The
Center for Exoplanets and Habitable Worlds is supported by the
Pennsylvania State University and the Eberly College of Science. 

This research has made extensive
use of the SIMBAD database, operated at CDS (Strasbourg, France) and
NASA's Astrophysics Data System Bibliographic Services.

\clearpage

\clearpage



\begin{deluxetable}{lllll}
\tablecaption{Stellar parameters of program stars. \label{Tab:Table0}}
\tablewidth{0pt}
\tablehead{\colhead{Parameter}         & BD+49~828              & HD~95127               & HD 216536 }
\startdata
$V$ [mag]                              & $9.38 \pm 0.02$        & $8.15 \pm 0.01$        & $9.23 \pm 0.02$ \\
Spectral type                          & K0                     & K0                     & K0 \\
$\pi$[mas]                             & ---                    & $3.06 \pm 0.99$        & ---   \\
\hline
$\Teff \left[\mathrm{K}\right]$        & $4943 \pm 30$          & $4218 \pm 69$          & $4639 \pm 45$ \\
$\log g$                               & $2.85 \pm 0.09$        & $1.78\pm 0.3$        & $2.36 \pm 0.21$ \\
${\rm [Fe/H]}$                         & $-0.19 \pm 0.06$       & $-0.18 \pm 0.05$        & $-0.17 \pm 0.09$ \\
$\vrotsini \left[\kmps\right]$         & $1.7 \pm 0.8$          & $2.6 \pm 0.7$          & $2.6 \pm 0.5$ \\
\hline
${\rm log}\,L_{\star}/L_{\odot}$       & $1.47 \pm 0.13$        & $2.28 \pm 0.38$        & $1.80 \pm 0.21$ \\
$M_{\star}/\Msun$                      & $1.52 \pm 0.22$        & $1.20 \pm 0.22$        & $1.36 \pm 0.38$ \\
$R_{\star}/\Rsun$                      & $7.6 \pm 1.3$          & $20 \pm 9$          & $12.5 \pm 4.0$ \\
$\log\left(\mathrm{Age}\right)$ [yr]   & $9.37 \pm 18$          & $9.74 \pm 0.27$        & $9.58 \pm 0.33$ \\
\hline
\hline
$\psini$ [days]                        & $226 \pm 114$          & {$389 \pm 204$}          & $243 \pm 91$ \\
$K_\mathrm{osc} \left[\mps\right]$     & $4.5^{+3.1}_{-1.8}$    & {$37^{+79}_{-25}$}       & $11^{+15}_{-6}$ \\
$P_\mathrm{osc}$ [days]                & $0.13^{+0.08}_{-0.05}$ & {$1.08^{+1.73}_{-0.81}$} & $0.39^{+0.56}_{-0.25}$    
\enddata
\end{deluxetable}



\begin{deluxetable}{lrrrr}
\tablecaption{Relative radial velocities and bisector span of BD+49 828 \label{Tab:Table1}}
\tablewidth{0pt}
\tablehead{ \colhead{Epoch (MJD)} & RV (m s$^{-1}$)  & $\sigma_{\textrm{RV}}$ (m s$^{-1}$) &
             BS (m s$^{-1}$)  & $\sigma_{\textrm{BS}}$ (m s$^{-1}$)}
\startdata
53225.44499 & -20.0 & 44.1 & 79.1 & 10.8 \\
53299.22458 & -2.9 & 15.2 & -9.8 & 26.6 \\
53593.40169 &  4.0 & 14.1 & -39.1 & 24.1 \\
53608.39577 &  0.5 & 13.5 & -42.3 & 21.6 \\
53610.36711 & 15.2 & 13.6 & -38.8 & 19.5 \\
54071.31499 & 17.8 & 13.5 & -12.2 & 15.9 \\
54087.06326 & 18.9 & 13.5 & -1.4 & 16.6 \\
54101.25992 & 17.0 & 13.7 & 34.6 & 16.1 \\
54114.21772 & -5.8 & 13.4 & 10.6 & 18.4 \\
54130.16924 & 24.2 & 13.6 & 90.2 & 15.0 \\
54148.11459 & -3.4 & 13.3 & 40.7 & 17.4 \\
54449.07380 &  4.6 & 13.7 &  7.6 & 22.5 \\
54457.27139 & 22.4 & 13.8 & 13.3 & 20.3 \\
54519.10534 & 34.1 & 13.7 & 58.9 & 17.5 \\
54731.27897 & -3.5 & 14.4 & 22.9 & 23.7 \\
54756.21112 & 10.5 & 14.2 & 12.4 & 22.3 \\
54781.17038 & 19.5 & 14.6 & 34.9 & 19.1 \\
55048.43124 & -27.9 & 14.3 & 23.6 & 10.3 \\
55076.35465 &  0.2 & 15.3 & -51.9 & 10.6 \\
55079.34718 & -9.3 & 13.5 & -6.3 & 10.6 \\
55107.26189 &  8.7 & 14.0 & 15.0 & 10.7 \\
55110.26035 &  4.0 & 13.7 &  6.3 & 10.9 \\
55179.07981 & 15.4 & 14.5 & 17.9 & 22.3 \\
55206.22428 & -25.1 & 13.7 &  4.2 & 15.1 \\
55245.12414 &  2.8 & 14.6 & 74.9 & 22.4 \\
55247.11719 & -9.5 & 14.1 & 70.3 & 21.0 \\
55443.34090 & -35.4 & 15.0 &  7.6 & 32.0 \\
55478.25636 & -39.1 & 13.7 & 41.8 & 16.5 \\
55501.18840 & -26.4 & 13.8 & 39.8 & 19.6 \\
55518.16037 & -5.7 & 13.5 &  0.6 & 17.0 \\
55775.45399 & -11.8 & 13.1 & 16.5 & 16.9 \\
55791.37892 & -10.6 & 13.5 & 37.5 & 19.9 \\
55799.38224 & -18.2 & 13.1 & 10.7 & 17.8 \\
55825.30463 & -9.6 & 13.6 & -3.5 & 15.6 \\
55859.21333 &  2.3 & 13.7 & 14.8 & 13.2 \\
55931.23168 & -10.6 & 13.6 & 31.3 & 13.7 \\
55973.12528 &  5.1 & 13.3 & 40.2 & 15.8 \\
56214.24030 &  7.0 & 13.6 &  7.7 & 20.6 \\
56257.13616 &  3.9 & 13.6 & 17.7 & 15.1 \\
56324.16275 & 31.9 & 18.5 & -35.1 & 51.9 \\
56347.08323 &  4.5 & 13.6 & 54.5 & 18.0 \\
56359.08653 &  7.8 & 14.2 & 46.3 & 23.6 
\enddata
\end{deluxetable}

\begin{deluxetable}{lrrrr}
\tablecaption{Relative radial velocities and bisector span of HD~95127.\label{tab-rv_bs_measurements-HD_95127}}
\tablewidth{0pt}
\tablehead{ \colhead{Epoch (MJD)} & RV (m s$^{-1}$)  & $\sigma_{\rm{RV}}$ (m s$^{-1}$) & BS (m s$^{-1}$)  & $\sigma_{\rm{BS}}$ (m s$^{-1}$)}
\startdata
 53024.52304   &         38.3   &          7.1   &         23.1   &         17.8\\
 53370.35059   &        -82.1   &          9.1   &         12.3   &         24.0\\
 53867.22760   &       -138.3   &          5.5   &         34.6   &         13.3\\
 54048.49149   &         47.8   &          5.7   &          5.4   &         18.4\\
 54129.28005   &        137.1   &          6.6   &         19.8   &         17.9\\
 54153.43199   &        120.2   &          6.1   &         -4.5   &         16.9\\
 54165.40075   &         32.2   &          5.7   &         58.9   &         15.8\\
 54198.31076   &         37.0   &          4.3   &         32.6   &         13.3\\
 54212.26155   &         91.8   &          5.9   &         16.1   &         19.7\\
 54214.26450   &         87.1   &          4.8   &          2.9   &         18.3\\
 54233.22538   &        -35.9   &          4.6   &         21.3   &         14.0\\
 54249.17289   &        -43.3   &          5.2   &          6.8   &         15.9\\
 54249.17781   &        -46.4   &          4.3   &         -8.3   &         12.1\\
 54413.47777   &        -88.4   &          5.3   &         55.3   &         17.1\\
 54439.43738   &        -51.3   &          6.4   &         57.8   &         18.7\\
 54498.47970   &         11.5   &          7.1   &          8.8   &         21.1\\
 54532.18166   &        107.8   &          7.1   &         55.3   &         18.4\\
 54567.33124   &        150.4   &          6.9   &         22.1   &         18.5\\
 54568.09186   &         99.1   &          6.8   &         63.9   &         18.5\\
 54580.27156   &         75.8   &          5.8   &         31.9   &         15.6\\
 54618.14668   &        112.7   &          6.4   &         41.6   &         17.5\\
 54811.41721   &        -40.9   &          6.4   &        -19.2   &         19.5\\
 54842.33315   &        -35.1   &          5.9   &         17.0   &         18.4\\
 54867.24900   &        -34.6   &          6.1   &         44.6   &         18.0\\
 55173.39559   &        -78.0   &          5.7   &         23.6   &         16.2\\
 55197.34812   &         -4.3   &          4.8   &         13.8   &         13.3\\
 55221.51289   &       -108.7   &          4.3   &         -3.2   &         14.6\\
 55245.45360   &       -216.0   &          6.4   &         -3.8   &         14.8\\
 55358.13133   &       -182.0   &          5.4   &         29.3   &         13.3\\
 55512.49483   &         56.7   &          6.7   &         46.8   &         17.5\\
 55544.40921   &        188.8   &          6.8   &         30.3   &         17.6\\
 55574.31534   &         59.4   &          5.4   &         38.6   &         13.4\\
 55577.32536   &         64.5   &          6.6   &         18.5   &         18.7\\
 55607.45310   &         89.3   &          6.3   &         23.7   &         15.0\\
 55637.15901   &         37.6   &          5.9   &         65.4   &         16.1\\
 55652.10131   &         56.9   &          4.9   &         45.4   &         11.6\\
 55691.21517   &        -24.8   &          6.0   &         -0.7   &         12.8\\
 55718.16090   &        -89.4   &          6.1   &          0.4   &         14.1\\
 55870.50090   &       -141.0   &          8.5   &          1.5   &         28.7\\
 55917.38294   &          3.1   &          5.2   &        -17.7   &         13.3\\
 55953.51961   &        -42.6   &          4.9   &         33.6   &         12.8
\enddata
\end{deluxetable}

\begin{deluxetable}{lrrrr}
\tablecaption{Relative radial velocities and bisector span of HD 216536 \label{Tab:Table3}}
\tablewidth{0pt}
\tablehead{ \colhead{Epoch (MJD)} & RV (m s$^{-1}$)  & $\sigma_{\textrm{RV}}$ (m s$^{-1}$) &
             BS (m s$^{-1}$)  & $\sigma_{\textrm{BS}}$ (m s$^{-1}$)}
\startdata
53544.38601 & -43.1 & 24.5 & -39.4 & 17.4 \\
54365.34050 & 24.9 & 23.7 & 64.5 & 14.5 \\
54658.32385 & 32.8 & 25.1 & 45.0 & 30.8 \\
55370.39329 & 53.7 & 23.6 & -5.7 & 15.4 \\
55470.35085 & -63.9 & 23.8 & -12.0 & 13.1 \\
55515.18324 & -44.2 & 24.6 & -41.7 & 12.4 \\
55518.18400 & -30.3 & 24.0 &  8.9 & 14.9 \\
55728.36984 &  8.8 & 24.0 & 26.5 & 17.3 \\
55733.37812 & 41.2 & 23.7 & 27.8 & 13.5 \\
55740.37428 &  4.0 & 24.3 & -5.8 & 17.8 \\
55753.32927 & -13.2 & 23.8 & 13.9 & 14.2 \\
55758.31630 & -25.3 & 23.9 & -4.8 & 16.5 \\
55763.29701 & -21.3 & 24.2 &  6.2 & 15.0 \\
55768.28240 & -54.0 & 23.8 & 20.3 & 13.1 \\
55775.27116 & -53.2 & 23.7 & 19.6 & 11.5 \\
55780.22843 & -68.7 & 23.9 &  5.6 & 16.9 \\
55793.43019 & -57.6 & 23.8 & 34.3 & 14.8 \\
55801.20164 & -25.2 & 23.8 & -2.5 & 13.3 \\
55810.16312 & -21.8 & 23.8 & -15.7 & 13.9 \\
55816.12715 & -40.3 & 24.2 &  1.0 & 17.9 \\
55817.38332 & -8.9 & 23.8 & -11.3 & 14.9 \\
55822.36050 &  4.5 & 23.7 & 27.5 & 13.7 \\
55825.14133 & 34.6 & 23.9 & 33.1 & 14.6 \\
55826.14602 & 49.4 & 23.8 & 29.2 & 12.4 \\
55829.10961 & 58.6 & 23.7 & -13.9 & 13.6 \\
55829.34425 & 65.0 & 23.7 & -7.1 & 12.1 \\
55830.11034 & 75.2 & 23.7 & -3.2 & 13.3 \\
55840.31604 & 66.9 & 23.8 & 42.7 & 11.3 \\
55849.31763 &  5.4 & 23.9 & 25.4 & 11.8 \\
55858.05332 & 38.7 & 23.9 & 46.9 & 12.3 \\
55888.21126 & -23.4 & 23.7 & 41.2 & 12.1 \\
55903.14193 & -44.3 & 24.0 & 17.1 & 17.8 \\
55924.08639 & -5.2 & 23.9 &  9.7 & 14.1 \\
55932.06380 & -49.4 & 23.7 & 15.1 & 10.8 \\
56073.43822 & -42.4 & 23.8 & 34.7 & 17.5 \\
56080.43409 & -21.2 & 23.7 & 16.6 & 12.4 \\
56179.39309 & 25.8 & 23.8 & 23.8 & 16.0 \\
56196.11293 & 11.0 & 23.7 &  7.5 & 13.4 \\
56210.06818 & -64.3 & 23.7 &  4.5 & 11.7 \\
56228.25215 & -56.5 & 23.9 & -7.7 & 11.8 \\
56261.14362 & 13.6 & 23.6 & 17.5 & 11.6 \\
56262.16696 & 34.9 & 23.6 & 12.1 & 12.9 \\
56282.10282 & 28.3 & 23.9 & 25.6 & 15.9 \\
56287.10367 &  4.6 & 24.3 & 40.7 & 21.4 \\
56288.09887 & 12.8 & 23.8 & 35.0 & 11.6 \\
56291.08204 & 44.2 & 23.8 & 17.2 & 12.4 \\
56300.06651 & 38.7 & 23.8 & 31.8 & 13.2 
\enddata
\end{deluxetable}

\begin{deluxetable}{lccc}
\tablecaption{Orbital parameters of BD+49 828 b,  HD~95127 b and HD 21653  b.\label{Tab:Table5}}
\tablewidth{0pt}
\tablehead{ \colhead{Parameter} & BD+49 828 b & HD~95127 b& HD 21653 b }
\startdata
$p$ (DAYS)                              & $2590^{+300}_{-180}$   & $482^{+5}_{-5}$        & $148.6^{+0.7}_{-0.7}$\\
$T_0$ (MJD)                             & $55470^{+200}_{-170}$  & $53200^{+50}_{-50}$    & $53587^{+11}_{-11}$\\
$K$ ($\textrm{m s}^{-1}$)               & $18.8^{+6.2}_{-2.0}$   & $116^{+16}_{-9}$       & $50^{+8}_{-4}$ \\
$e$                                     & $0.35^{+0.24}_{-0.10}$ & $0.11^{+0.15}_{-0.06}$ & $0.38^{+0.12}_{-0.10}$ \\
$\omega$ (deg)                          & $170^{+32}_{-30}$      & $40^{+37}_{-40}$       & $270^{+21}_{-20}$\\
$m_2\sin i$ ($\textrm{M}_{\textrm{J}}$) & $1.6^{+0.4}_{-0.2}$    & $5.01^{+0.61}_{-0.44}$ & $1.47^{+0.20}_{-0.12}$\\
$a$ (AU)                                & $4.2^{+0.32}_{-0.2}$   & $1.28^{+0.01}_{-0.01}$ & $0.609^{+0.002}_{-0.002}$\\
$V_0$ ($\textrm{m s}^{-1}$)             & $1.1^{+1.4}_{-0.5}$    & $-10.5^{+3.1}_{-2.3}$  & $-4.9^{+0.7}_{-2.1}$ \\
$\sqrt{\chi_\nu^2}$                     & $1.35$                 & $1.14$                 & $1.29$ \\
$\sigma_{\textrm{RV}}\; (\mps)$         & $11.6$                 & $50.9$                 & $23.0$ \\
jitter $(\mps)$                         & $4.44^{+0.35}_{-1.00}$ & $47.5$               & $17.9$ \\
$N_{\textrm{obs}}$                      & $42$                   & $41$                   & $47$ 
\enddata
\end{deluxetable}

\begin{deluxetable}{llllllllll}
\tablecaption{Basic parameters of planets around stars at various stages of stellar evolution (from exoplanets.eu). Mean and median values as well as dispersion are presented for every set of data. See text for details.  \label{Tab:Table6}}
\tablewidth{0pt}
\tablehead{ 
\colhead{Parameter} & M$_{\star}$ range  & dwarfs & & subgiants&  & giants & & bright giants & \\ 
\hline
\colhead{} &  & mean & $\sigma$ &  mean & $\sigma$  &  mean & $\sigma$  &  mean & $\sigma$ \\
\colhead{} &  & median & &  median &   &  median &   & median &  \\ }
\startdata
N$_{planet}$   &1-2 M$_{\odot}$ & 267 	& & 73 & & 23 & & 9 & \\
                      &   all                    &   458 	& &  77   & & 41    & & 14    & \\
 		     &1-2 M$_{\odot}$(RV) & 138 	& & 60 & & 22 & & 9 & \\
                      &   all(RV)                    &   236 	& &  64   & & 40    & & 14    & \\
                      &                           &        	& &      & &     & &     & \\   
log(g)                      			&   all                    &   4.378 	& 0.009	&  3.540	& 0.037	& 2.574    & 0.039 	& 1.594   	& 0.077\\
                      			&                           &   4.377	& 		&   3.394   & 		&  2.594   & 		&  1.663   	& \\   
           				&1-2 M$_{\odot}$ & 4.285 	&0.008 	& 3.533 	& 0.037	& 2.533 	& 0.054 	& 1.529 	& 0.100\\
                      			&                           & 4.282 	&  		&  3.386    & 		& 2.501   	& 		&  1.521   	& \\    

                      			&   all(RV)                    &   4.392 	& 0.013	&  3.485	& 0.038	& 2.564    & 0.038 	& 1.594   	& 0.077\\
                      			&                           &   4.392	& 		&   3.369   & 		&  2.582   & 		&  1.663   	& \\   
                      			&  1-2 M$_{\odot}$(RV)                    &   4.296 	& 0.012	&  3.473	& 0.038	& 2.514    & 0.053 	& 1.529   	& 0.100\\
                      			&                           &   4.300	& 		&   3.350   & 		&  2.501   & 		&  1.521   	& \\   

&&&&&&&&&\\

M$_{\star}/M_{\odot}$                      			&   all                    &  1.039	& 0.011	&  1.428	& 0.027	& 1.870    & 0.090 	& 1.464   	& 0.120\\
                      			&                           &  1.031	& 		&   1.450   & 		&  1.900   & 		&  1.395   & \\  
					&1-2 M$_{\odot}$ & 1.179 	&0.009 	& 1.453	& 0.025	& 1.499 	& 0.060 	& 1.457 	& 0.084\\
                      			&                           & 1.140 	&  		&  1.470    & 		& 1.500   	& 		&  1.400	& \\                      
                      			&   all(RV)                   &  0.997	& 0.015	&  1.446	& 0.031	& 1.885    & 0.091 	& 1.464   	& 0.120\\
                      			&                           &  1.030	& 		&   1.475   & 		&  1.900   & 		&  1.395   & \\    
                      			&   1-2 M$_{\odot}$(RV)                   &  1.140	& 0.010	&  1.478	& 0.028	& 1.507    & 0.062 	& 1.457   	& 0.084\\
                      			&                           &  1.100	& 		&   1.480  & 		&  1.500   & 		&  1.400   & \\   

&&&&&&&&&\\

$[$Fe/H$]$                      	&   all                    &  0.071	& 0.010	& 0.086	& 0.022	& -0.032    & 0.032 	& -0.247   	& 0.058\\
                      			&                           &  0.090	& 		& 0.120   & 		& -0.052   & 		&  -0.255   & \\   
					&1-2 M$_{\odot}$ & 0.114 	&0.011 	& 0.099	& 0.020	& -0.012 	& 0.045 	& -0.222 	& 0.055\\
                      			&                           & 0.140 	&  		& 0.140    & 		& -0.030   	& 		&  -0.260	& \\                      
                      			&   all(RV)                    &  0.102	& 0.014	& 0.093	& 0.026	& -0.036    & 0.033 	& -0.247   	& 0.058\\
                      			&                           &  0.140	& 		& 0.140   & 		& -0.056   & 		&  -0.255   & \\ 
                      			&   1-2 M$_{\odot}$(RV)                    &  0.178	& 0.012	& 0.109	& 0.023	& -0.018    & 0.046 	& -0.222   	& 0.055\\
                      			&                           &  0.210	& 		& 0.140   & 		& -0.060   & 		&  -0.260   & \\ 

&&&&&&&&&\\
               
m$_{P}$sin{i}/m$_{J}$      &   all                    &  2.046	& 0.147	& 2.702	& 0.399	& 6.391    & 0.987 	& 7.798   	& 1.186\\
                      			&                           &  0.955	& 		& 1.800   & 		& 4.500   & 		&  6.300   & \\   
					&1-2 M$_{\odot}$ & 2.459 	&0.212 	& 2.705	& 0.416	& 4.557 	& 0.592 	& 7.688 	& 0.854\\
                      			&                           & 1.308 	&  		& 1.800    & 		& 3.200   	& 		&  7.800	& \\    
                      			&   all(RV)                    &  2.250	& 0.200	& 2.956	& 0.467	& 6.533    & 1.001 	& 7.798   	& 1.186\\
                      			&                           &  1.135	& 		& 1.850   & 		& 4.900   & 		&  6.300   & \\ 			   			
                      			&   1-2 M$_{\odot}$(RV)                    &  2.532	& 0.256	& 2.977	& 0.492	& 4.731    & 0.592 	& 7.688   	& 0.854\\
                      			&                           &  1.658	& 		& 1.850   & 		& 3.260   & 		&  7.800   & \\ 

&&&&&&&&&\\
                    
e                      			&   all                    & 0.162 	& 0.010	& 0.185	& 0.018	& 0.187    & 0.031 	& 0.220   	& 0.048\\
                      			&                           & 0.072 	& 		& 0.160   & 		& 0.110   & 		&  0.167   & \\
					&1-2 M$_{\odot}$ & 0.165 	&0.013 	& 0.178	& 0.018	& 0.177 	& 0.045 	& 0.207 	& 0.068\\
                      			&                           & 0.070 	&  		& 0.157    & 		& 0.090   	& 		& 0.140	& \\  	
                      			&  all(RV)                   & 0.244 	& 0.014	& 0.212	& 0.019	& 0.190    & 0.032 	& 0.220   	& 0.048\\
                      			&                           & 0.201 	& 		& 0.185   & 		& 0.111   & 		&  0.167   & \\  					
                      			&   1-2 M$_{\odot}$(RV)                    & 0.260 	& 0.018	& 0.205	& 0.019	& 0.182    & 0.047 	& 0.207   	& 0.068\\
                      			&                           & 0.225 	& 		& 0.179   & 		& 0.095   & 		&  0.140   & \\  

&&&&&&&&&\\
                  
a/AU                      		&   all                    & 1.116 	& 0.193	& 1.331	& 0.111	& 1.476    & 0.172 	& 1.487   	& 0.085\\
                      			&                           & 0.094 	& 		& 1.310   & 		& 1.200   & 		&  1.470   & \\  
					&1-2 M$_{\odot}$ & 1.442 	&0.324 	& 1.358	& 0.115	& 1.119	& 0.113 	& 1.517 	& 0.119\\
                      			&                           & 0.095 	&  		& 1.310    & 		& 1.160   	& 		& 1.540	& \\  
                      			&   all(RV)                    & 1.506 	& 0.108	& 1.568	& 0.112	& 1.511    & 0.173 	& 1.487   	& 0.085\\
                      			&                           & 0.983 	& 		& 1.450   & 		& 1.235   & 		&  1.470   & \\ 			  			
                      			&  1-2 M$_{\odot}$(RV)                   & 1.725 	& 0.149	& 1.617	& 0.115	& 1.166    & 0.107 	& 1.517   	& 0.118\\
                      			&                           & 1.188 	& 		& 1.515   & 		& 1.180   & 		&  1.540   & \\ 

&&&&&&&&&\\
 %
%
&&&&&&&&&\\
\enddata
\end{deluxetable}


\begin{deluxetable}{lcccccccc}
\tablecaption{Parsons correlation coefficients between stellar hosts mass and total mass of their planetary systems. \label{Tab:Table7} 
Upper values: all systems for which at least masses are available; lower values: systems with all data listed in Table 6 available.
All correlation coefficients are smaller than respective critical values.}
\tablecolumns{4}
\tablewidth{0pt}
\tablehead{\colhead{group}  & \multicolumn{4}{c}{M$_{PS}$ vs. M$_{*}$} & \multicolumn{4}{c}{N$_{systems}$} \\
\colhead{}   & \colhead{1-2M$\odot$} & \colhead{all} & \colhead{1-2M$\odot$ (RV)} & \colhead{all (RV)} & \colhead{1-2M$\odot$} & \colhead{all} & \colhead{1-2M$\odot$ (RV)} & \colhead{all (RV)} \\}
\startdata
\multirow{2}{*}{dwarfs}  & 0.222 & 0.132 & 0.194 & 0.234 & 256 & 432 & 100 & 172 \\
   & 0.137 & 0.172 & 0.202 & 0.223 & 220 & 367 & 99  & 167 \\

\multirow{2}{*}{subgiants} & 0.128 & 0.133 & 0.110 & 0.109 & 68  & 73  & 55  & 59  \\
   & 0.119 & 0.113 & 0.095 & 0.099 & 64  & 68  & 53  & 57  \\

\multirow{2}{*}{giants}& 0.101 & 0.409 & 0.057 & 0.400 & 21  & 40  & 20  & 39  \\
   & 0.182 & 0.480 & 0.144 & 0.469 & 20  & 37  & 19  & 36  \\

\multirow{2}{*}{bright giants} & 0.114 & 0.604 & 0.114 & 0.604 & 9   & 16  & 9   & 16  \\
   & 0.114 & 0.606 & 0.114 & 0.606 & 9   & 14  & 9   & 14  \\

\multirow{2}{*}{all}   & 0.204 & 0.234 & 0.164 & 0.402 & 354 & 567\footnote{Including 6 hosts with log(g)$>$5.} & 184 & 286 \\
   & 0.181 & 0.380 & 0.172 & 0.407 & 313 & 486 & 180 & 274 \\
\enddata
\end{deluxetable}

\clearpage


\clearpage

\clearpage

\clearpage

\clearpage

\clearpage

\clearpage

\clearpage

\clearpage

\clearpage

\clearpage

\clearpage

\end{document}